\documentclass[preprint2]{aastex}

\usepackage{graphicx}
\usepackage{epstopdf}
\usepackage[caption=false,...]{subfig}
\usepackage{longtable}
\usepackage{lscape}

\shorttitle{VVV survey: galaxies behind the Galactic plane}
\shortauthors{Am\^{o}res et al.}

\begin{document}


\title{Galaxies behind the Galactic plane: First results and
perspectives from the VVV Survey}


\author{E. B. Am\^{o}res\altaffilmark{1,2}}
\author{L. Sodr\'{e}\altaffilmark{3}}
\author{D. Minniti\altaffilmark{4,5,6}}
\author{M. V. Alonso\altaffilmark{7,8}}
\author{N. Padilla\altaffilmark{4,9}}
\author{S. Gurovich\altaffilmark{7,8}}
\author{V. Arsenijevic\altaffilmark{1}}
\author{E. J. Tollerud\altaffilmark{10}}
\author{A. Rodr\'{i}guez-Ardila\altaffilmark{2}}
\author{J. D\'{i}az Tello\altaffilmark{7}}
\author{P. W. Lucas\altaffilmark{11}}


\altaffiltext{1}{FCUL, Campo Grande, Edificio C5, $1749-016$
Lisboa, Portugal.}

\altaffiltext{2}{Laborat\'{o}rio Nacional de Astrof\'{i}sica /
MCTI, Rua Estados Unidos 154, Itajub\'{a}$-$MG, 37504-364,
Brazil.}

\altaffiltext{3}{Instituto de Astronomia, Geof. e Ci\^encias
Atmosf\'ericas da USP, Cidade Universit\'aria 05508-900 S\~ao
Paulo SP, Brazil}

\altaffiltext{4}{Departamento de Astronom\'{i}a y Astrof\'{i}sica,
Pontificia Universidad Cat\'{o}lica de Chile, Santiago, Chile.}

\altaffiltext{5}{Vatican Observatory, Vatican City State
$V-00120$, Italy.}

\altaffiltext{6}{Department of Astrophysical Sciences, Princeton
University, Princeton, NJ 08544-1001, USA.}

\altaffiltext{7}{Instituto de Astronom\'{i}a Te\'{o}rica y
Experimental (IATE-CONICET), Laprida 922 X5000BGR C\'{o}rdoba,
Argentina.}

\altaffiltext{8}{Observatorio Astron\'{o}mico de C\'{o}rdoba,
Laprida 854, C\'{o}rdoba, Argentina.}

\altaffiltext{9}{Centro de Astro-Ingenier\'{i}a, Pontificia
Universidad Cat\'{o}lica de Chile, Av. Vicu\~{n}a Mackenna 4860,
Santiago, Chile.}

\altaffiltext{10}{Center for Cosmology, Department of Physics and
Astronomy, University of California Irvine, Irvine, CA 92697,
USA.}

\altaffiltext{11}{Centre for Astrophysics Research, University of
Hertfordshire, Hatfield, UK.}


\begin{abstract}
Vista Variables in The Via L\'{a}ctea (VVV) is an ESO variability
survey that is performing observations in near infrared bands
(ZYJHK$_s$) towards the Galactic bulge and part of the disk with
the completeness limits at least 3 mag deeper than 2MASS. In the
present work, we searched in the VVV survey data for background
galaxies near the Galactic plane using ZYJHK$_s$ photometry that
covers 1.636 square degrees. We identified 204 new galaxy
candidates by analyzing colors, sizes, and visual inspection of
multi-band (ZYJHK$_s$) images. The galaxy candidates colors were
also compared with the predicted ones by star counts models
considering a more realistic extinction model at the same
completeness limits observed by VVV. A comparison of the galaxy
candidates with the expected one by Milennium simulations is also
presented. Our results increase the number density of known
galaxies behind the Milky Way by more than one order of magnitude.
A catalog with galaxy properties including ellipticity, Petrosian
radii and ZYJHK$_s$ magnitudes is provided, as well as comparisons
of the results with other surveys of galaxies towards Galactic
plane.
\end{abstract}



\section{Introduction}

Large surveys of Galactic structure offer an unique opportunity to
study extra-galactic sources such as background galaxies and
quasars. However, studies of such sources and their structures in
the Galactic plane are limited due to the high density of
foreground sources. Moreover, one of the most severe obstacle at
low Galactic latitude is the interstellar extinction that can
reach values of up 35 magnitudes in A$_V$ (e.g. Neckel \& Klare,
1980, Am\^{o}res \& L\'{e}pine 2005, hereafter AL05) that clearly
making surveys at optical wavelengths unfeasible for this type of
study.

Pioneering works studying the complex structures behind the Milky
Way (MW) Galactic plane had been started by Kraan$-$Korteweg et
al. (1995), Fairall et al. (1998), Woudt et al. (1998) and Tonry
et al. (2001). They were motivated by the Great Attractor (GA)
model suggested by Lynden$-$Bell et al. (1988), a mass
concentration in the direction of Hydra$-$Centaurus supercluster
on l=307$^\circ$ and b=9$^\circ$. This concentration would be
responsible for the observed peculiar motions of elliptical
galaxies in the nearby Universe (Dressler et al. 1991).

Even though some of the early predictions from such
reconstructions indicated the presence of a GA behind the MW, data
from the X-ray cluster survey in the Zone of Avoidance (ZoA,
Ebeling et al. 2000), showed that such a mass would be about one
order of magnitude smaller (Kocevski \& Ebeling, 2005, Kocevski et
al., 2007, see also Radburn-Smith et al., 2006).

In addition, the group led by Woudt \& Kraan-Korteweg have
performed a deep optical galaxy search behind the southern MW
towards Hydra/Antlia (Kraan-Korteweg 2000) and the GA region
(Woudt \& Kraan-Korteweg 2001). These two works presented over
11,000 previously unrecorded galaxies with average observed major
axis diameters equal to 0.2 arcmin. Later, they also identified
342 IRAS sources. Woudt et al. (2004) using Meudon-ESO Fibre
Object Spectrograph (MEFOS) presented redshifts for 764 sources
towards the ZoA. Much progress has been made in the last years in
the study of galaxies behind the Galactic plane with radio surveys
as well as those in the near-mid and far-infrared and X-rays,
among others.

The Two Micron All Sky Survey (2MASS, Skrutskie et al. 1997)
covers the whole sky in J, H and K passbands. Jarrett et al.
(2000a) presented a basic algorithm to detect, characterize and
extract extended (e.g., extragalactic) sources using 2MASS data.
Jarrett et al. (2000b) performed an internal completeness and
reliability analysis for a sample of 2MASS data, consisting of
7,000 sources in a region of 1,000 square degrees of area,
including galaxies and Galactic nebulae from the W51 giant
molecular cloud complex. We note that several of these newly
discovered galaxies found in photometric surveys are the closest
known galaxies outside the MW (eg., Tollerud et al. 2008).

Hurt et al. (2000) using 2MASS data presented a late-type spiral
galaxy (Sc-Sd) located at $\sim 11$ Mpc, ($\ell,b$)
$=(236.82^{o},-1.86^{o})$, with an angular extent of 6.3$'$ in the
near infrared. Skrutskie et al. (2006) published 'The 2MASS
Extended Source Catalog' that contains 1,647,599 sources that are
extended with respect to the instantaneous PSF, such as galaxies
and Galactic nebulae.

As pointed out by Jarrett et al. (2000b) near-infrared surveys
have the advantage of being sensitive to most types of galaxies,
including gas-poor spheroidals. In this sense, Nagayama et al.
(2004) presented a deep NIR survey covering a region of 36
arcmin$^{2}$ centered on the radio galaxy PKS 1343-601, providing
57 galaxy candidates. Based on DENIS data, Paturel et al. (2005)
published a catalogue containing IJK photometry for 753,153
sources, including galaxies and galaxies to be confirmed.
Schr\"{o}eder et al. (2007) investigated a 30 square-degree area
around the radio-bright galaxy PKS 1343-601 ($\ell,b$)
$=(309.7^{o},+1.8^{o})$ using the DENIS survey. They found 83
galaxies and 39 galaxy candidates.

Due to the Galaxy transparency at longer wavelengths it is very
useful to observe galaxies using HI (21 cm line). Kraan-Korteweg
et al. (2002) identified 66 galaxies using the Parkes 64 m radio
telescope. Ryan-Weber et al. (2002) and Koribalski et al. (2004)
identified approximately 1,000 objects using HI Parkes All-Sky
Survey (HIPASS). And, in a further work, Schr\"{o}eder et al.
(2009), using 64 m Parkes radio telescope, made observations in HI
for 314 candidates, detecting 162 of them. Radburn-Smith et al.
(2006) using the 2dF on the AAT measured 3053 redshifts in the
GA/SSC region and Van Driel et al. (2009) performed HI
observations towards 132 sources identifying 16 2MASS galaxies in
their sample.

Recently, Jarrett et al. (2007) using the Spitzer Legacy program
GLIMPSE identified two galaxies towards a high extinction zone
($\ell,b$) $= (317^{o},-0.5^{o}$) with A$_V \sim 15.0$ mag. They
complemented their observations with HI data from ATCA survey.
Marleau et al. (2009) using GLIMPSE and MIPSGAL mid-infrared
surveys found twenty-five galaxy candidates towards $\ell \sim$ 47
and 55$^{o}$.

As pointed out by Woudt et al. (2004), deep optical searches, near
infrared all-sky surveys (2MASS and DENIS), HI all-sky surveys and
X-ray surveys, have also all resulted in the detection of voids,
clusters and superclusters at low Galactic latitude (for a review,
see Kraan-Korteweg \& Lahav 2000 and references therein).

The goal of this work is to identify galaxies behind the Galactic
plane in the Vista Variables in the Via L\'{a}ctea, hereafter
VVV\footnote{http://www.vvvsurvey.org} (Minniti, D., et al. 2010
and Saito, et al. 2012), a deep near IR survey of the MW bulge and
part of the disk. Due to the specific area of interest, two main
difficulties are: i-) highly crowded fields, ii-) the effect of
interstellar extinction.

The interstellar dust absorption (and reddening), does affect the
observed color distribution of galaxies, and hence causes some
galaxies to drop out in the surveys (a selection bias). Minimizing
such biases is necessary for an accurate estimation of galaxy
number statistics and also for the accurate determination and
distribution of Galactic extinction.

This is the first of a series of papers that unveil galaxies in
the Galactic plane using deep VVV near infrared photometry. The
results presented here were obtained by visual inspection of the
false-color and multi-band images, color-color and color
histogram, size distribution analysis for the objects in a region
towards $\ell=$ 298.3558, $b=$-1.6497 (called the VVV d003
region). Following this procedure, we have found around 200 galaxy
candidates. We present results for only one tile to establish
methods that can be used in future works. Furthermore, this single
tile will allow us to identify galaxies that have no counterpart
in previous surveys.

This paper is organized as follows. Section 2 presents the main
characteristics of the VVV data. Section 3 presents an overview of
the properties of the d003 region, such as interstellar extinction
and stellar distribution. Our method to identify galaxies is
presented in Section 4 along with comparison to data observed by
other catalogs towards a region with 25 square degrees centered at
d003. A discussion and comparison of the photometry for both
stellar and sources that we identified as galaxies as well as a
comparison with mock catalogues of galaxies are given in Section
5. Finally, Section 6 presents the conclusions and the final
remarks.

\section[]{VVV data}

VVV is an ESO public variability survey with the 4-m Vista
telescope at Cerro Paranal. It consists of observations in the
near infrared bands (Z, Y, J, H, and K$_s$) towards the Galactic
bulge and part of the disk, covering in total an area of 562
square degrees. The VVV is in its third year; we already have full
disk and bulge coverage in J, H and K$_s$ and near completion in Y
and Z (see Saito et al. 2012 for a detailed description of the
survey status). The completeness limit of the observations in the
5 infrared bands is 21.6 mag in Z, 20.9 in Y, 20.6 in J, 19.0 in H
and 20.0 mag in K$_s$.

As pointed by Saito et al. (2012), VIRCAM  is equipped with 16
Raytheon VIRGO $2048  \times 2048$ pixels$^2$  HgCdTe science
detectors, with  $0\farcs 339$  average pixel scale. Each
individual detector therefore covers $\sim 694 \times
694$~arcsec$^2$ on the sky. The  achieved image quality (including
seeing)  is better  than $\sim 0\farcs6$ on axis. The image
quality distortions are up to about 10\% across the  wide field of
view.

The data were reduced using the Cambridge Astronomy Survey Unit
(hereafter CASU) pipeline\footnote{http://casu.ast.cam.ac.uk/}.
The provided data comprises 62 attributes, including positions and
magnitudes for each detected source. The positions are converted
into J2000 equatorial coordinates (RA,DEC) and the sources in
different filters are matched (with radius search equal to 1
arcsec) in a final catalogue. Fluxes are transformed into
magnitudes calibrated taking into account the zero points and
atmospheric extinction.

Typical   values  for  the astrometric  accuracy  are $\sim25$~mas
for a  $K_{\rm  s}=15.0$~mag source  and $\sim175$~mas  for
$K_{\rm  s}=18.0$~mag (Saito et al. 2012).

For this study we have selected from VVV database, the equatorial
coordinates, aperture magnitudes (within an aperture of 1.0 times
the core radius) and errors in all filters, a classification flag,
a parameter called `statistic`, which is a measure of the
stellarity of an object and the Petrosian radii (Yasuda, et al.
2001) in all filters. The classification flag gives an indication
of the most probable classification of a source: -2 if it is
compact (probably stellar), -1 if it is stellar, 0 if it is
classified as a noise peak, and 1 if it is extended, non-stellar.
Details about the data reduction algorithms are presented in Irwin
et al. (1994).

\section{VVV d003 tile}

The d003 tile analyzed in this work was observed in the ZYJHK$_s$
bands and is centered at (J2000) RA=12:09:17.18 and
DEC=-64:08:46.7. The tile was selected by its relatively low and
uniform extinction with complete observations using the 5 filters
(ZYJHK$_s$) and multi-epoch K$_s$ band observations.

This field covers 1.636 square degrees ($\Delta\ell \sim$ 1.475
degrees and $\Delta b \sim$ 1.109 degrees). It contains a huge
number of detected objects, including extended, stellar and
compact sources: 1,095,483; 1,018,374; 1,233,111; 1,302,302; and
1,160,528 in the ZYJHK$_s$ filters, respectively. Figure 1 shows a
combined JHK$_s$ false-color image of part of this region,
depicting some background galaxies. One of the main goals of this
work is to characterize these objects.

From the point of view of Galactic structure, the line of sight
towards this tile is not far from the tangential direction to
Carina$'s$ spiral arm at $\ell \sim$ $285^{o}$ to $290^{o}$ (e.g.,
Bronfman 1992, Russeil 2003 and references therein).

\subsection{The stellar and interstellar extinction distributions}

Star counts are very useful to model observed data allowing
several studies.  These include the distribution of the Galactic
objects in color space, the relation between colors and distance
that can be used to infer extinction and to correct observed
colors for interstellar extinction effects. This is useful in
order to obtain an accurate estimate of the number of Galactic
sources for a given region and to correlate it with the observed
extra-galactic sources.

Star counts in the Galactic plane have been modeled by Am\^{o}res
\& Robin (2012, in preparation) with the Besan\c{c}on Galaxy Model
(BGM, Robin et al. 2003) by fitting Galactic parameters such as
spiral arms, disc scale-length, the presence of the disk warp and
flare, to stellar counts obtained by the 2MASS survey in the
Galactic plane, covering an area equal to 300 square degrees and
allowing to derive new parameters and description for the spiral
arms, warp and flare distribution in the MW.

In the present work, we have used the version of the BGM with the
Marshall et al. (2006) extinction model to simulate star counts
towards d003. Our simulation strategy consisted in separating the
tile into small regions of around $0.25^{o}$ by $0.25^{o}$ square
degrees. This tiling is adequate due to the irregular distribution
of the extinction and stellar content. Before simulating star
counts, we computed the completeness limits for each filter J, H
and K$_s$ as well as their photometric error, modelled by using an
exponential law (Bertin \& Arnouts, 1996). The results obtained
from star counts simulations will be presented in Section 4.
Because BGM does not include the Z and Y filters, we used for them
the TRI-LEGAL (Girardi et al. 2005) model.

We have also analyzed the distribution of interstellar extinction
in the field of the d003 tile based on the maps\footnote{Just for
the record, note that the resolution of Froebrich et al. (2005)
and Dobashi et al. (2005) maps are 2 and 4 arcmin, respectively.
For both maps we have used the four neighbourhood extinction
values to calculate the extinction provided directly in A$_V$
units.} provided by Froebrich et al. (2005), Dobashi et al. (2005)
and the AL05 models. The median values for A$_V$ within the tile
using them are 2.27 $\pm$ 0.23, 2.66 $\pm$ 0.71, 3.00 $\pm$ 0.79,
respectively.

For a more detailed discussion about the interstellar extinction
towards d003 tile, see Table 4 of Saito et al. (2012). Also,
previous works searching for galaxies in the Galactic plane found
A$_V$ ranging from 3.0 mag to 15.0 mag, as discussed by
Kraan-Korteweg et al. (2000), Woudt \& Kraan-Korteweg (2001),
Nagayama et al. (2004), and Jarrett (2007). Our values for the
tile d003 ($\sim$ 3 mags) are at lower end of this range. For a
short and recent review on interstellar extinction, see Robin
(2009) and references therein.

\section{VVV d003 tile: Visually Identifying Galaxies}

The main goal of this paper is to carry out the first detection,
in the literature, of new galaxies in the VVV survey. We started
with the d003 tile as a first step to establish the basic
procedure to search for these objects. This section presents a
brief description of the data obtained by other surveys. The
expected number of galaxies predicted by Millennium simulation is
presented in Section 5.1.

\subsection{Previous Works}

It is worth verifying whether there are galaxies in the d003
region already identified by other surveys. We first concentrated
on the catalog published by Skrutskie et al. (2006) based on 2MASS
data that found no galaxies in the d003 tile area. Indeed, only
342 sources were classified as galaxies by 2MASS in the complete
VVV region, corresponding to only 0.65 galaxies$/$square degrees.

Table 1 shows the sources observed by 2MASS with visual
verification of source (\textit{vc}) equal to -2, -1, 1 and 2,
which corresponds to unknown, no verification, galaxy,
non$-$extended (e.g., star, double, triple, artifact),
respectively for three different regions, i.e, VVV bulge and disk
and an area equal to 40$^{o}$ square degrees centered in d003 tile
center. Although there are 65 objects (Table 1) identified as
galaxy by 2MASS in this area, no object classified as galaxy in
d003 tile is found. Even if we relax the \textit{vc} flag
constraint to get all non-stellar sources in our analysis, we note
that no extended 2MASS sources were detected in the d003 tile.

We have also checked that no galaxies were found towards the d003
tile in another studies of galaxies behind the Galactic plane
provided by Paturel et al. (2005), Koribalski et al. (2004), Woudt
et al. (2001). Figure 2 shows DIRBE/COBE (Freudenreich 1998)
emission map at 2.2 $\mu$m. As can be seen by comparing the number
of galaxies, we may expect that an analysis of all regions
observed by the VVV survey will largely increase the number of
galaxies detected behind the Galactic plane. We therefore conclude
that, up to our knowledge, no galaxies in the d003 tile were
previously identified in that region.

\subsection{Analysis of the VVV d003 tile}

The Galactic plane is very rich in sources with many different
objects. They can be classified as point sources, mostly Galactic
stars, and extended images, including blended stars, background
galaxies, star formation regions, planetary nebulae, supernova
remnants, molecular clouds, etc. Because of this great variety of
objects, a logical and cautious first approach is also to confirm
the identification of sources obtained by VVV pipeline (galaxies
for the purposes of this investigation) by visual inspection.
Based on this information, we can later launch a full automatic
exploration of galaxies within the entire 562 square degrees area
covered by the VVV Survey.

As discussed earlier, the identification of galaxies near the
Galactic plane suffers from two major difficulties: high Galactic
extinction and dense stellar fields. Also, extended features in
the interstellar medium are an additional source of confusion.

Initially we have analyzed by visual inspection a false-color
image (in J, H, K$_s$) of the d003 tile using Aladin (Bonnarel et
al. 2000) in order to identify extended sources that we consider
to have galaxy features. A total of 473 sources were found in this
way and we can see some examples in Figure 1. Figure 3 also shows
twenty objects classified by us as galaxy candidates located at
the center of figures for several directions within d003 tile with
their number in our catalog.

By using the coordinates of the objects, we have performed a
cross-id search of 5 arcsec in the CASU database. As a next step,
we made a superposition of these sources in the false-color image
in order to assure that the sources found in this search
correspond to those obtained through visual inspection and if the
source has galaxy features. The resulting number of sources
obtained with this analysis is 300.

A more robust method for identifying galaxies in large surveys is
an analysis of properties that can be performed automatically.
These include colors, sizes, and ellipticity. In the next section
we present an analysis of these objects using these properties.

\section{Identification of d003 galaxies}

Color-color diagrams are useful tool for discriminate galaxies
from stars in an objective way, since they occupy well defined
regions in the diagrams. However, due to the high crowding in the
Galactic plane region, most of the extended sources could be, in
fact, double or multiple stars.

Covey et al. (2007) traced the location of the main-sequence stars
in the ugrizJH$K_s$ color-color space with a sample containing
$\sim$ 600,000 point sources observed by 2MASS and SDSS (Abazajian
et al. 2009), providing fits of stellar color locus as a function
of $g-i$ color. Hewett et al. (2006) also presented synthetic
color-color diagrams for stars and galaxies in a study of UKIRT
infrared photometric system.

We have analyzed the colors of 300 galaxy candidates in the
ZYJHK$_s$ color-space by means of three color-color diagrams
(Figure 4) as well as J-K$_s$ and H-K$_s$ color histograms. We
have flagged objects as belonging to the stellar locus for both
objects located in the high star density region\footnote{we have
defined as high star density region the space in the color-color
diagrams in which the colors are red and yellow in the diagrams of
Figure 4.} of each color-color diagram and have colors  J-K$_s <
1.0$ or H-K$_s < 0.3$. Next, multi-band visual classification was
performed independently by six co-authors classifying the objects
as likely galaxy, possible galaxy (see classification below), or
to be discarded, e.g. seems to be a star, appearing star-like in
all bands: Z, Y, J, H and K$_s$.

We have classified the galaxy candidates in two different
categories. Type I sources (possible galaxies) have galaxy
features and are photometrically consistent with those expected
for galaxies. Type II (likely-galaxies) sources do not have clear
Galactic features, but are photometrically consistent with
galaxies and do not appear to have PSF-like shapes. For Type II,
additional verification is necessary.

Final classification is based on the objects 'color distribution
and the co-authors' visual inspection of the multi-band images.
For the objects outside the high star density regions in the
color-color diagrams, at least three co-authors must give the same
visual classification for an object to be classified as Type I or
II. For the objects at high stellar density, we had more caution,
requiring five authors to agree on the classification.

Figure 4 shows color-color density maps with the colors obtained
from BGM and TRI-LEGAL star counts models (both introduced in the
Section 4) over the entire d003 tile. The colors for galaxy
candidates are the observed ones in the VVV and for star counts
simulations take into account a more realistic extinction model.
We show objects classified as Type I (possible) as black crosses
and Type II (likely-galaxies) as red diamonds, revealing a
separation between stars and galaxies. In particular, Figure 4a
(J-H vs H-K$_s$ color-color diagram) shows significant separation
between the objects classified as both Type I and mostly of Type
II and the region of high stellar density.

There are some objects classified by us as galaxy-like that are
located in moderate density stellar regions, mostly for the
diagrams that involved Y and Z filters, which can also be
attributed to high interstellar extinction. However, most of these
objects show clear galaxy features, and their NIR colors are
outside the highest stellar density regions. Thus, we kept those
objects in our list with flag equal to '2' in Table 2.

As pointed out by Jarrett et al. (2000a), two effects conspire to
make galaxies appear "red" in the $1-2$ $\mu$m window: their light
is dominated by older and redder stellar populations (e.g., K and
M giants), and their redshifts tend to transfer additional stellar
light into the 2 $\mu$m window (for z $<$ 0.5), boosting the
K-band flux relative to the J-band flux (K-correction). They also
noticed that J-K color $\sim 1.0$ is a reasonable limit to
separate stars from galaxies. Another useful piece of information
from the color-color diagrams is that galaxies have H-K colors
larger than 0.3, redder than most of the stars.

The location of most of galaxy candidates in our color-color
diagrams (Figure 4) are also consistent with the synthetic ones
presented by Hewett et al. (2006) (see their Figure 3).

One of the main differences between our diagrams and those
presented by Hewett et al. (2006) is that the diagram represented
in Figure 4c is that ours shows a narrower range of colors for the
galaxies. This can be interpreted as an effect of interstellar
extinction (they considered A$_v = 0$) and mainly of redshift
since the galaxies in Hewet et al. span from 0 to 3.6.

In total, there are 72 objects classified as source Type I and 142
as source Type II. Figure 5 shows images of a sample of objects
classified as source Type I in the K$_s$ filter. Figure 6 shows
multi-band images for five possible galaxies. In total there that
are 44 objects detected in all of the five filters and 119 objects
with common data in J, H and K$_s$. Figure 7 shows a histogram
with counts of galaxy candidates. It can be seen that completeness
limit without extinction correction for this tile is around K$_s
=$ 16.5 mag.

Our final catalog of galaxy candidates contains magnitudes and
errors in five filters, the Petrosian radius, ellipticity and PA
for each object as well as our classification as presented in
Table 2 (that shows only part of the catalog). The full version is
available on-line. The ranking in this table is from 1 to 300
since it follows the object number as selected by-eye inspection
and with match in VVV database (Section 4.2).

Figure 8 shows the spatial distribution of the objects assigned as
source Type I (crosses) and source Type II (diamonds) for two
ranges of magnitudes K$_s \leq 16.5$ and K$_s > 16.5$ mag, denoted
with smaller and large symbols, respectively. The density counts
represent the star counts observed by VVV, corrected by the
effects of interstellar extinction using BGM. It can be seen that
in the region that corresponds to high stellar density the
detected galaxies counts decrease.

Another powerful tool to identify extended objects is through a
size versus magnitude diagram (Jarrett et al. 2000a). This
procedure allows us to discriminate between stars (point sources)
and extended sources in an objective way, since they occupy well
defined regions in this diagram. However, due to the high
crowdness in the region, most of the extended sources are, in
fact, double or multiple stars. Figure 9 shows three types of
sources in different colors: stars (point sources), Type I
(crosses) and II (red). In the present work only stellar located
within 5 arcsec of each galaxy candidate have been considered. For
both Type I and II, a Petrosian radius of $\sim$ 4.0 arcsec
separates most of stars from galaxy candidates.

\subsection{Mock catalogues}

We constructed two mock VVV galaxy catalogues extracted from the
Bower et al. (2007) semi-analytic galaxies that populate the
Millennium simulation (Springel et al., 2005). The Galactic
absorption from Marshall et al. (2006) is only included in one of
them. In order to construct the catalogues we place an observer in
the Millennium simulation box and include all galaxies with an
apparent, observer-frame K$_s < 16.5$ without and with extinction,
respectively covering the disc and bulge sections of the VVV
survey. The final number of galaxies in the catalogues is 61,332
and 34,854 without and with extinction, respectively.

Figure 10 shows the histogram of number of galaxies per VVV disc
tile of $1.5$ square degrees from the mocks (solid lines for the
catalogue with no Galactic extinction, dotted when the extinction
is taken into account and the average extinction for d003).  The
solid line shows the range of number of galaxies expected in tiles
due to fluctuations of the clustering of the background galaxies,
whereas the dotted line shows the combined effect of background
clustering and foreground extinction. On average, the lowest
numbers of galaxies in the catalogue with extinction correspond to
the highest extinctions (and vice-versa), however this is only a
statistical relation.  As can be seen in the previous section that
the number of galaxy candidates in disc tile d003 is typical for
the number of galaxies in the mocks for an average Galactic
extinction in the MW disc.

We can provide a coarse estimate of the total number of galaxies
to be found in the VVV Survey by inspecting the interstellar
models proposed by AL05. We count the number of VVV fields that
have similar or less total extinction than field d003, which is 40
in the disk and 140 in the bulge. Assuming the background galaxy
distribution throughout the disk and bulge is similar to that of
field d003, a wrong assumption since we are not taking into
account the crowdness distribution, but good as a first
approximation, we multiply the number of galaxies in field d003 by
60. The result allows us to predict that there would be of order
of 15,000 galaxies in the VVV Survey database.

\section{Conclusions}

Using ZYJHK$_s$ imaging, we identify 204 new galaxy candidates in
a region of the VVV covering 1.636 square degrees. This region is
near the Galactic plane, at ($\ell$,b) = (298.3558,-1.6497). We
have identified 72 objects as source Type I and 142 as source Type
II out of 204 sources.

We identify galaxy candidates both by visual inspection and by
excluding stars based on the colors, magnitudes, and sizes of our
objects. We compare our MW galaxy catalogue with those obtained by
other authors, over similar volume, and find that none\footnote{We
also have verified (February, 9 th) at NED/IPAC database and no
object classified as galaxy was found towards d003 tile.} of our
proposed galaxies have ever been detected in any other survey
mostly because they are too faint due to the effects of Galactic
extinction and crowdness.

The large obscuration in the Galactic plane is an obstacle for
detection of galaxies and large scale structures that exist in the
directions of the Galactic plane. Although with VVV we increase
the known number density of known galaxies in the d003 tile, we
will be able to increase our sensitivity ever further as we start
combining multi-epoch data as these become available.

The VVV survey brings considerable progress to the observation of
fainter sources behind the Galactic plane due to its resolution
and completeness limits. However, this is still a large number
compared with the expectations for the whole VVV Survey because
d003 is a relatively "clear" field. We have estimated that our
survey will detect about a billion point sources, and about 15,000
galaxies. The difficulty of this task is illustrated by the small
fraction of galaxy candidates found amongst so many stars of our
galaxy: only one out of about 5,000 objects is a galaxy, a small
fraction indeed (0.020\%).

In a forthcoming paper (Am\^{o}res et al., in preparation) it will
be presented an automatic method to separate galaxies from stars
in the Galactic plane and the detected galaxies by VVV survey for
both Galactic bulge and disk. With our method we should be able to
minimize the J-$K_s$ threshold problem that causes bluer galaxies
to drop out in surveys.

The observation of background galaxies and the related improvement
of photometric redshifts can also help to solve the open questions
left behind by matter density estimations performed using galaxy
peculiar velocities (e.g. POTENT) and X-ray surveys (CIZA) which
point out to the existence of relatively massive structures behind
the Milky Way. Thus, a map of projected number density of objects
will be a valuable addition to our understanding of nearby
large-scale structure to probe the existence of nearby Great
Attractor-like background objects. However, a crucial step is to
use d003 candidates as spectroscopic targets to really confirm the
nature of these candidates.  This will give us a statistical
estimates of the galaxy detection that is possible by using only
VVV data photometry.

\acknowledgments

We thank the anonymous referee for useful comments on the
manuscript. We gratefully acknowledge use of data from the ESO
Public Survey programme ID 179.B-2002 taken with the VISTA
telescope, and data products from the Cambridge Astronomical
Survey Unit, and funding from the FONDAP Center for Astrophysics
15010003, the BASAL CATA Center for Astrophysics and Associated
Technologies PFB-06, the MILENIO Milky Way Millennium Nucleus from
the Ministry of Economy´s ICM grant P07-021-F, and the FONDECYT
from CONICYT. Eduardo Am\^ores obtained financial support for this
work from Funda\c{c}\~{a}o para a Ci\^{e}ncia e Tecnologia (FCT)
under the grant SFRH/BPD/42239/2007. Laerte Sodr\'{e} Jr.
achnowledges the support of FAPESP and CNPq. ARA acknowledges CNPq
for partial support to this research through grant 308877/2009-8.
Eduardo Am\^ores thanks Dr Annie Robin and Dr Leo Girardi for
providing him their star count models. The mock background galaxy
catalog was constructed and analysed using the Geryon computer
cluster at the Centro de Astro-Ingeniería UC. EJT acknowledges a
GAANN fellowship for support. We also thank Ign\'{a}cio Toledo for
the tiles of the false-color image, Mike Read and Eduardo
Gonz\'{a}lez for help with VSA and CASU, respectively. This
publication makes use of data products from the Two Micron All Sky
Survey, which is a joint project of the University of
Massachusetts and the Infrared Processing and Analysis
Center/California Institute of Technology, funded by the National
Aeronautics and Space Administration and the National Science
Foundation. Some VVV tiles were made using Aladin sky atlas
(Bonnarel et al. 2000), SExtractor software (Bertin \& Arnouts
1996), and products from TERAPIX pipeline (Bertin et al. 2002).

\newpage
\clearpage

\begin{figure*}[h]
\centering \resizebox{12.0cm}{!}{\includegraphics{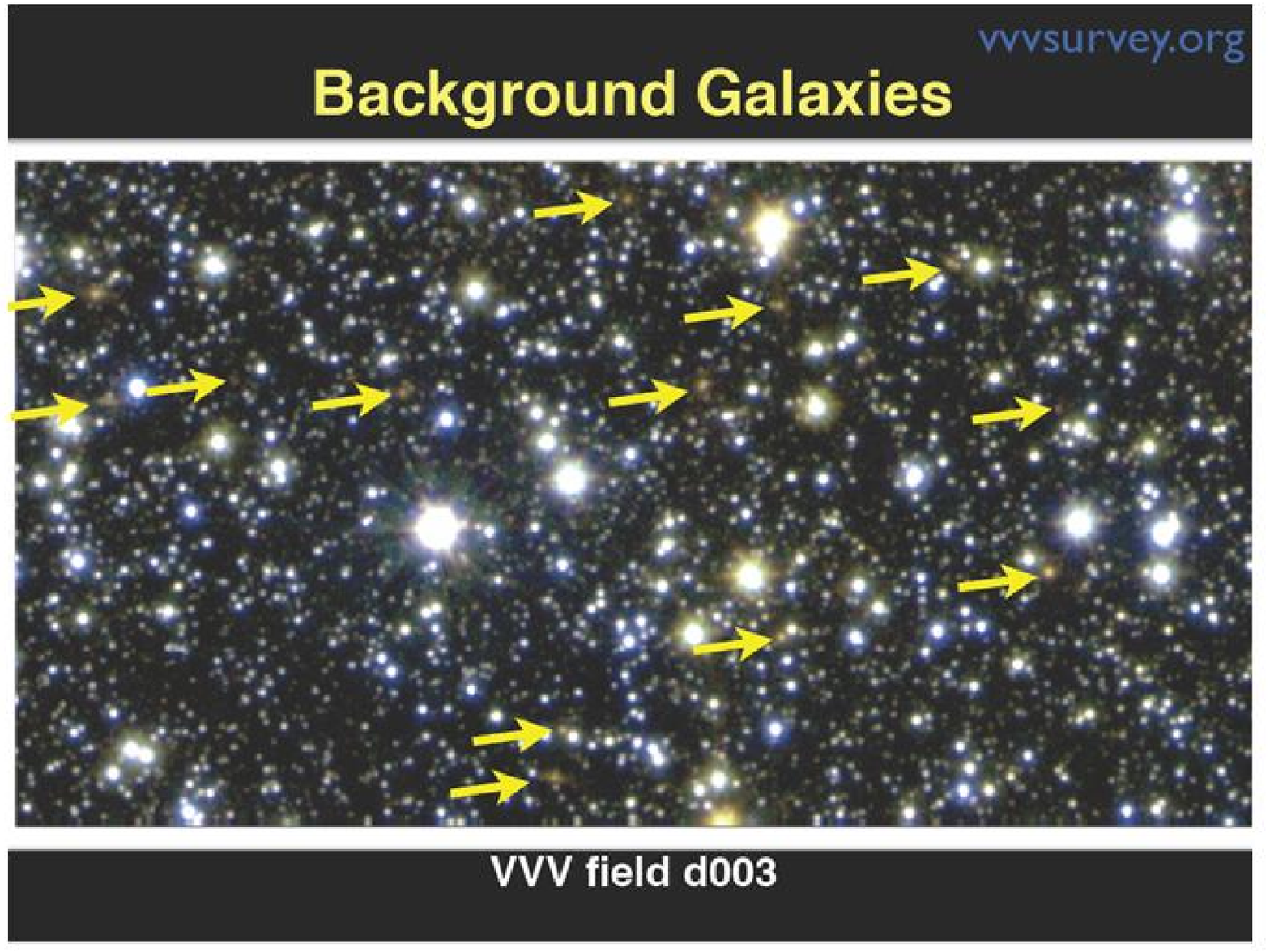}}
\caption{Cut out of a false-color image from J, H \& K$_s$
passbands in equatorial coordinates J2000 (RA,DEC) of the d003
tile, showing galaxy candidates.}
\end{figure*}

\clearpage

\begin{figure*}[h]
\centering
\resizebox{18.0cm}{!}{\includegraphics[angle=90,scale=1.0]{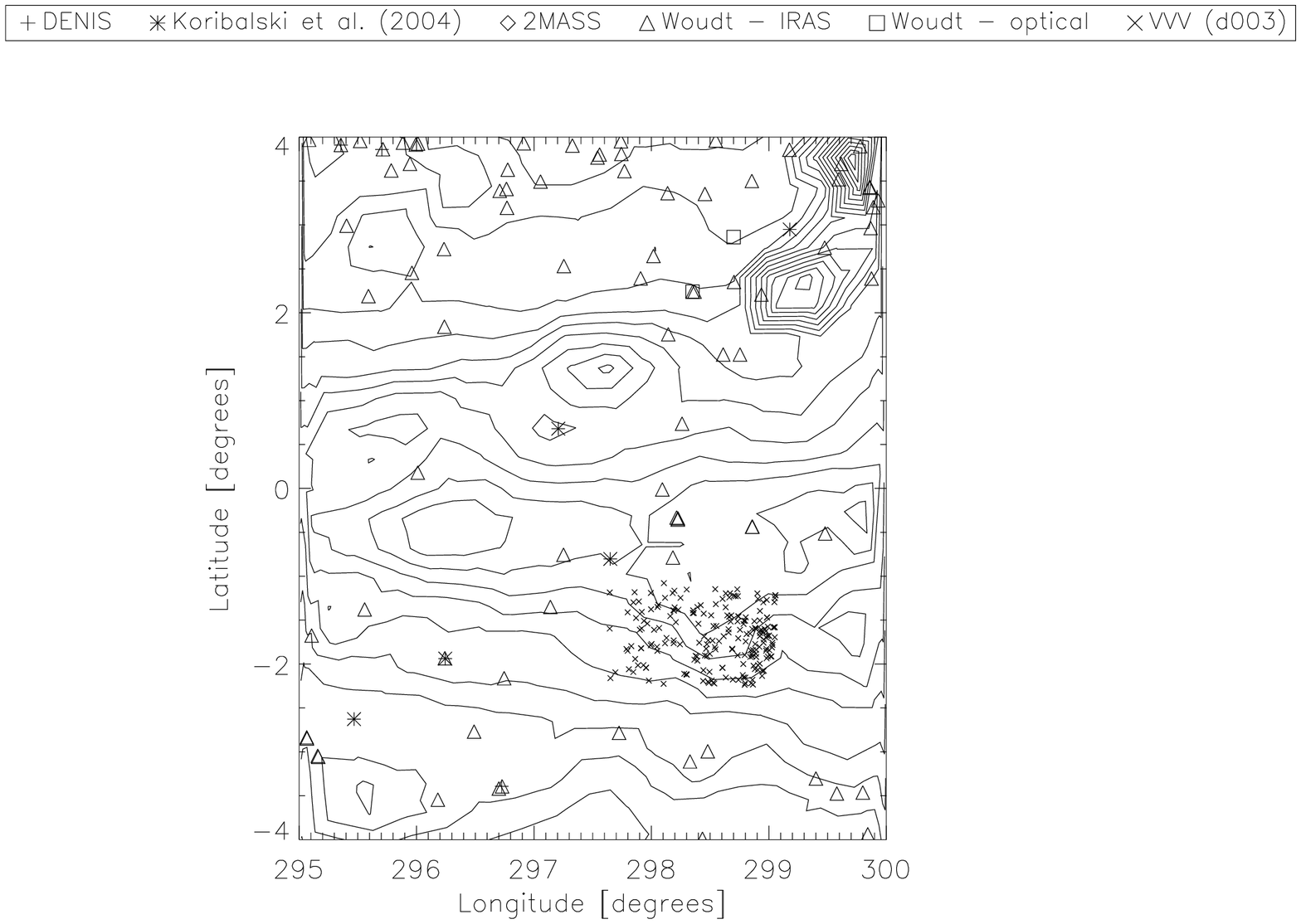}}
\caption{DIRBE/COBE emission map at 2.2 $\mu$m; each interval in
the contour represents 0.5 MJy; galaxies observed by other surveys
(see legend at the top) are large symbols; objects of the present
work are represented by crosses in d003 tile centered at
($\ell$,b) = (298.3558,-1.6497). References: DENIS (Paturel et
al., 2005); 2MASS (Skrutskie et al. 2006); Woudt et al. (2001).}
\end{figure*}

\clearpage

\begin{figure*}
\begin{minipage}{2.0\textwidth}
\begin{tabular}{cccc}
022 & 034 & 065 & 088 \\ \\ \\
\includegraphics[bb=0 0 868 543,width=3.0cm]{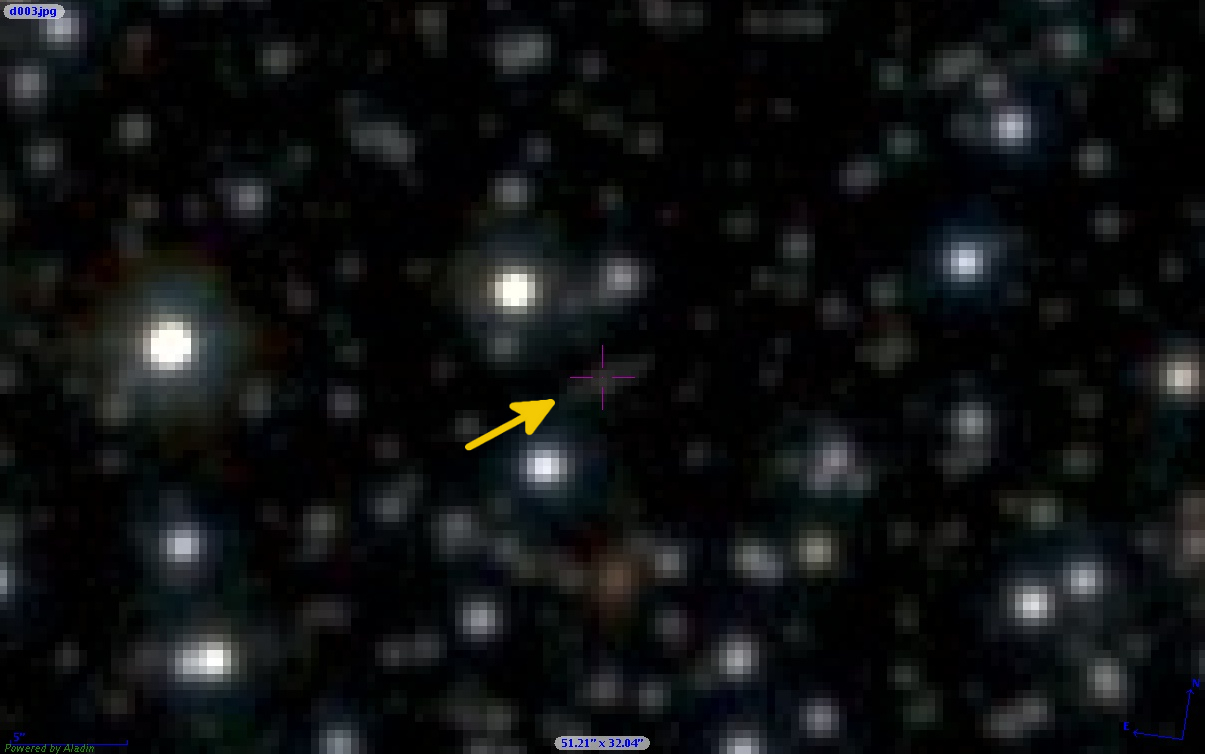} &
\includegraphics[bb=0 0 868 543,width=3.0cm]{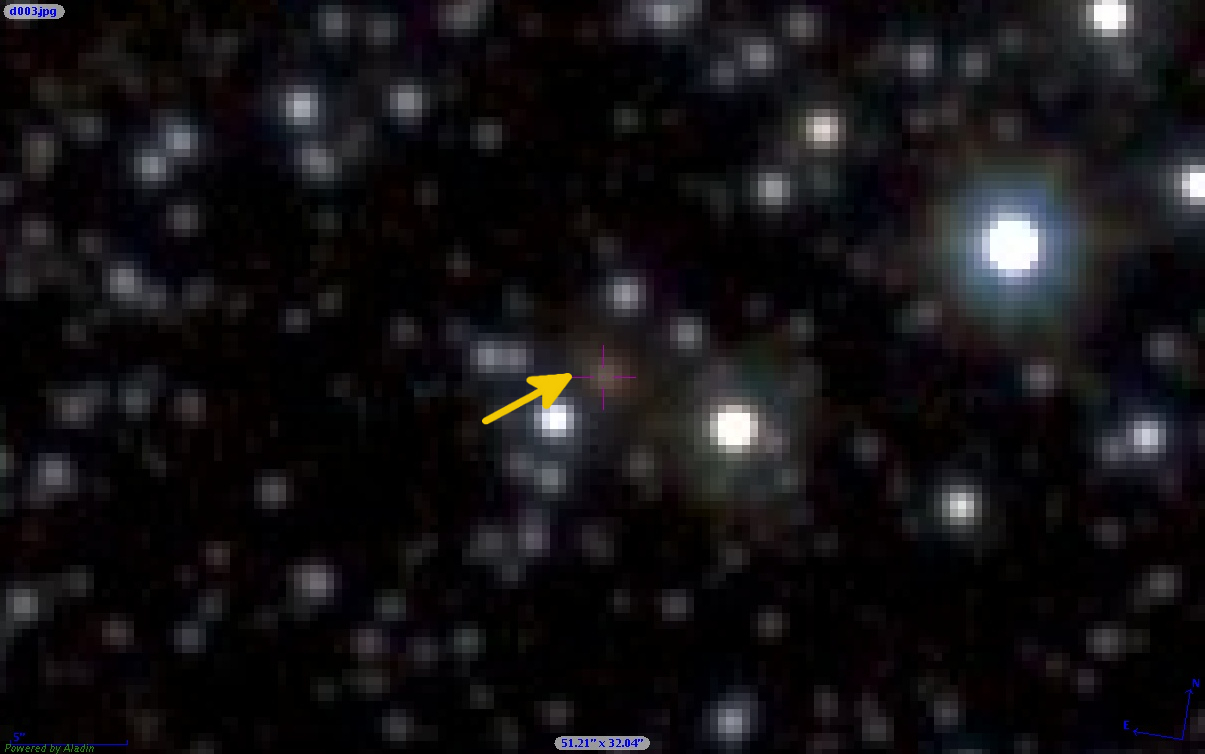} &
\includegraphics[bb=0 0 868 543,width=3.0cm]{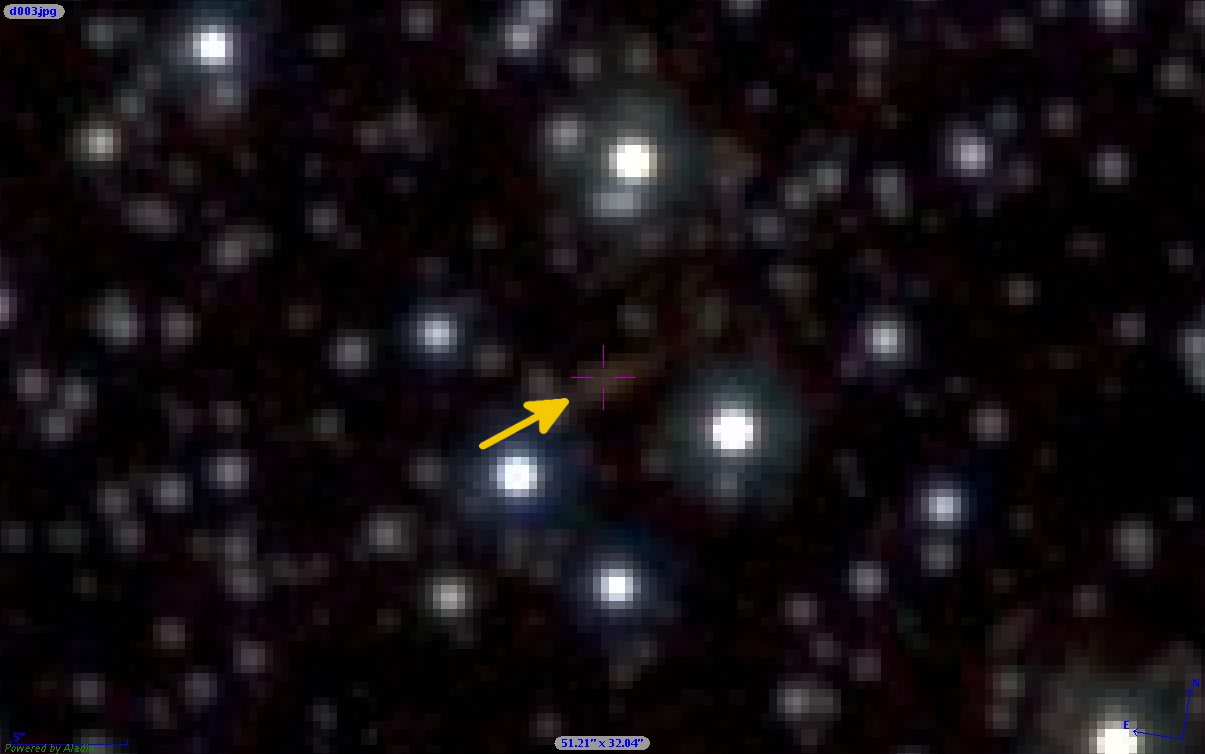} &
\includegraphics[bb=0 0 868 543,width=3.0cm]{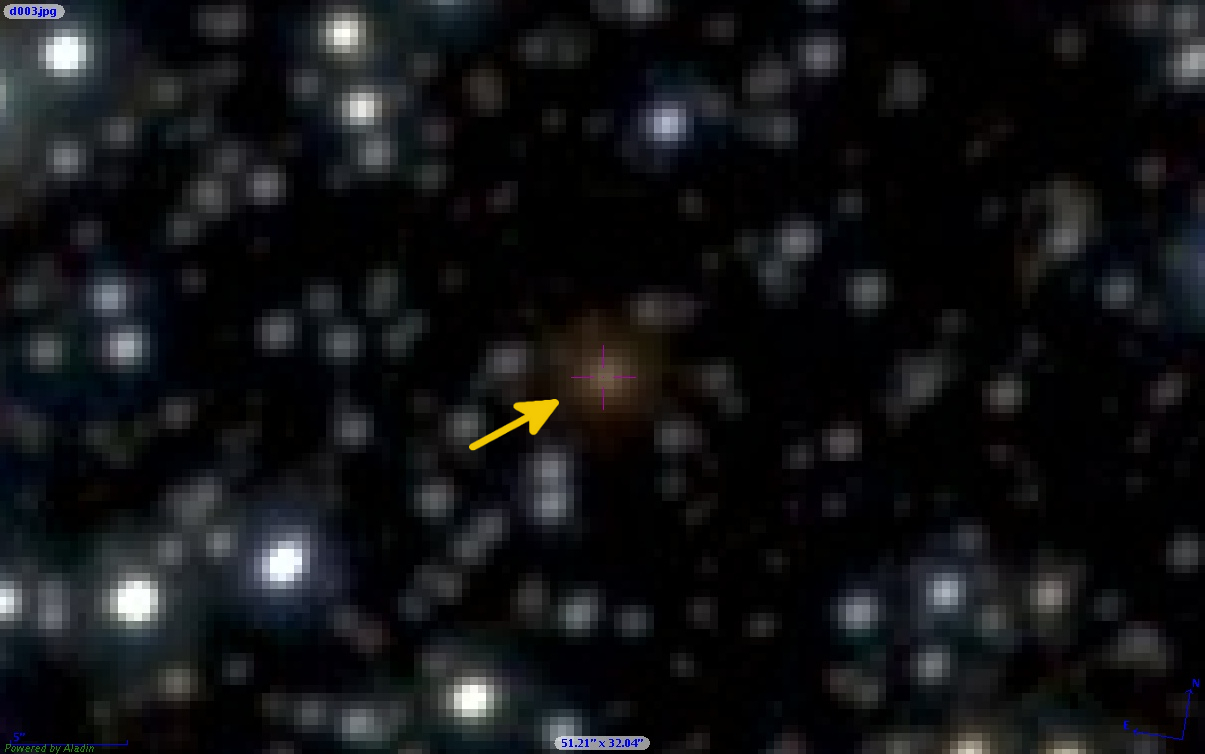} \\ 
089 & 135 & 148 & 171 \\ \\ \\
\includegraphics[bb=0 0 868 543,width=3.0cm]{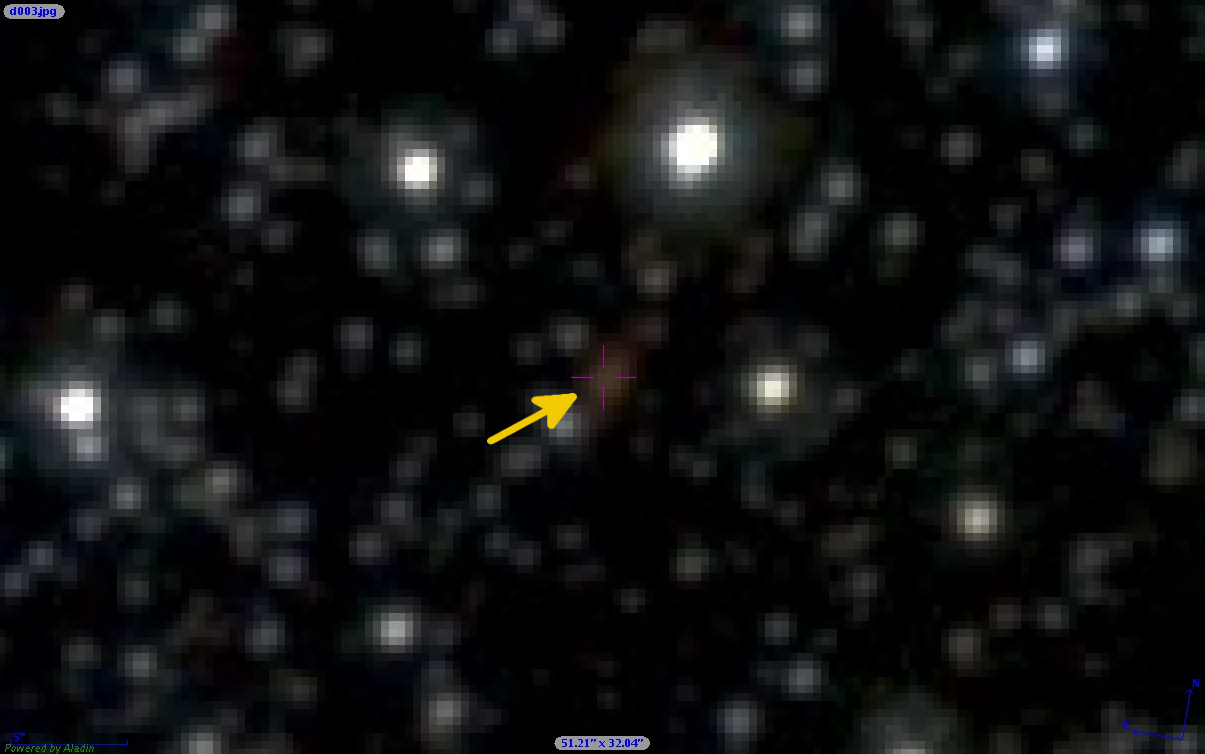} &
\includegraphics[bb=0 0 868 543,width=3.0cm]{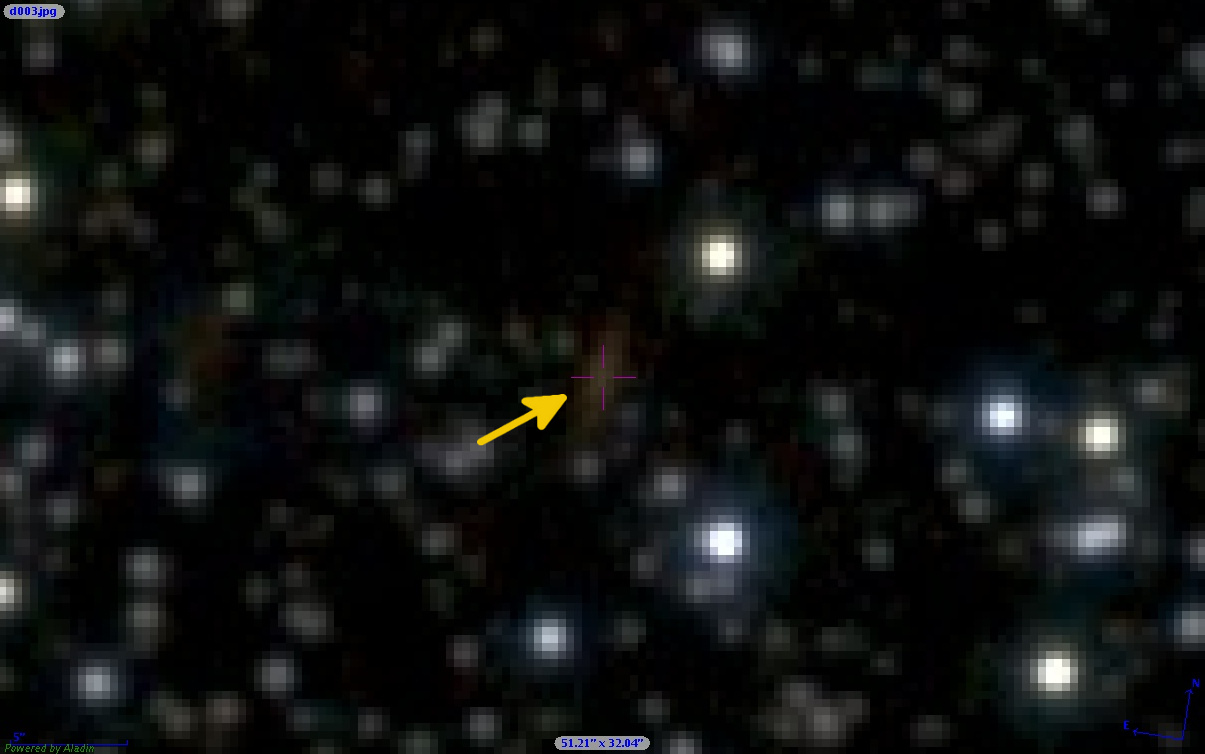} &
\includegraphics[bb=0 0 868 543,width=3.0cm]{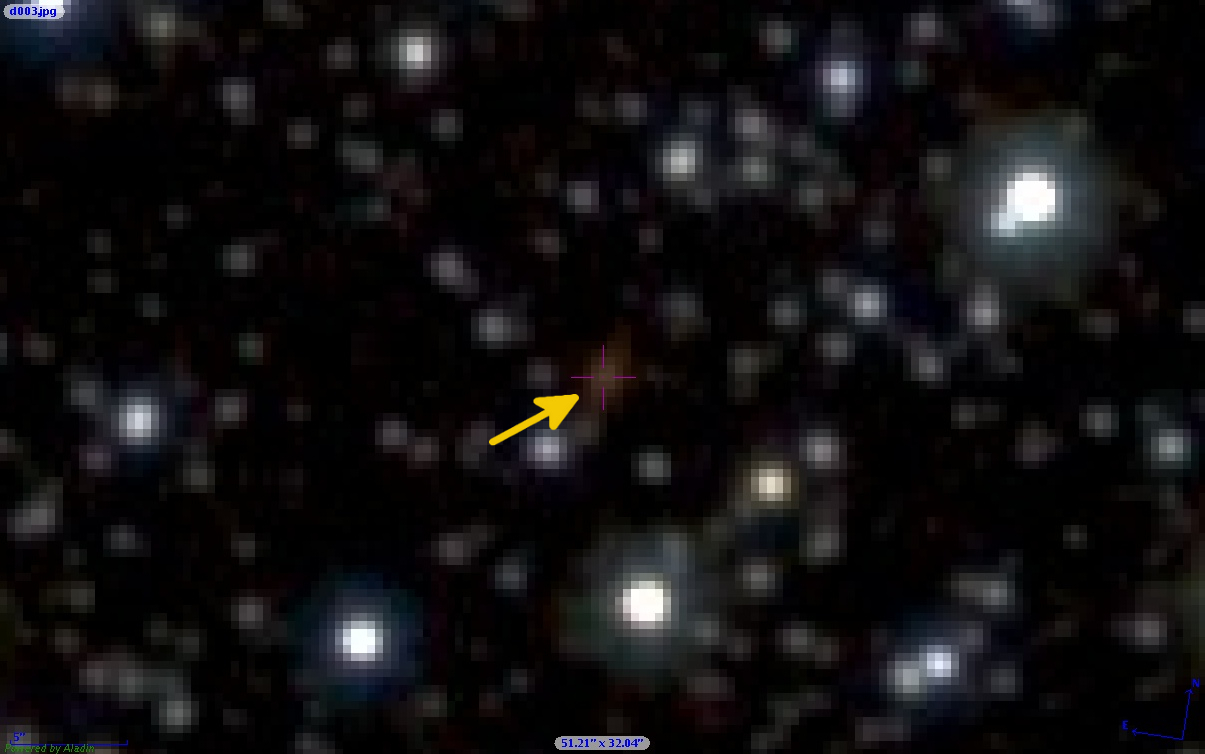} &
\includegraphics[bb=0 0 868 543,width=3.0cm]{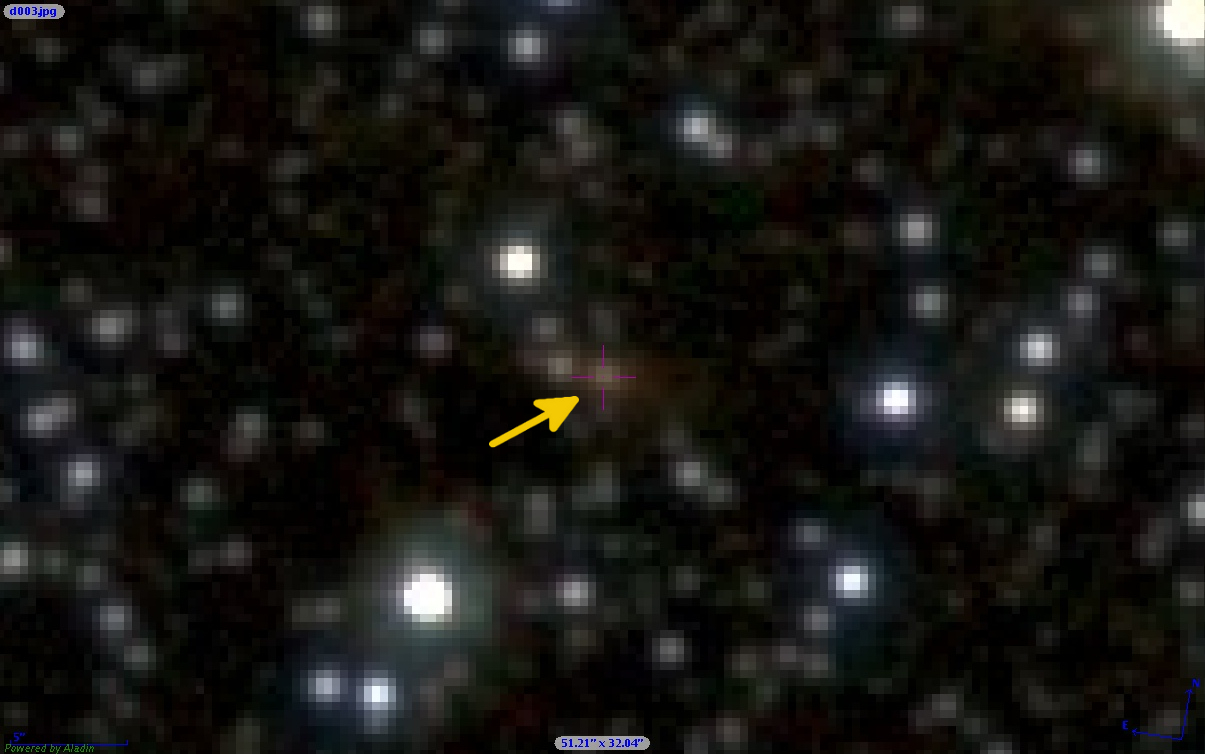} \\ 
175 & 191 & 196 & 201 \\ \\  \\
\includegraphics[bb=0 0 868 543,width=3.0cm]{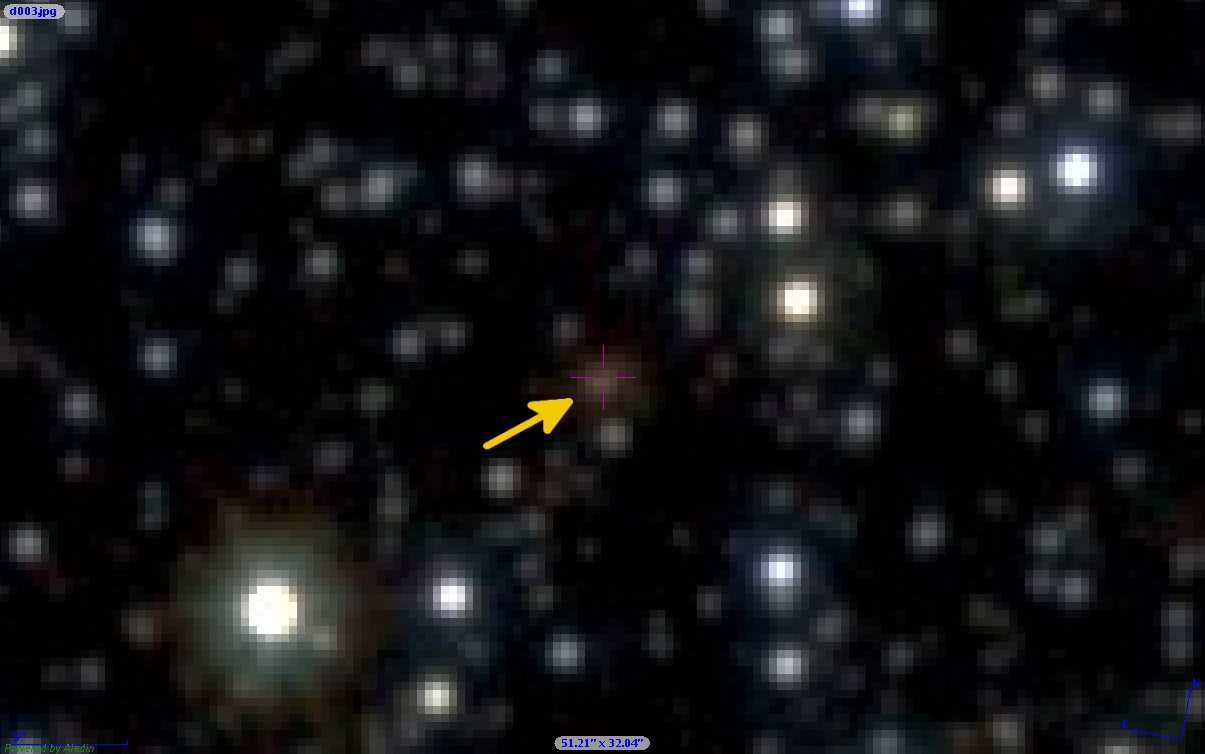} &
\includegraphics[bb=0 0 868 543,width=3.0cm]{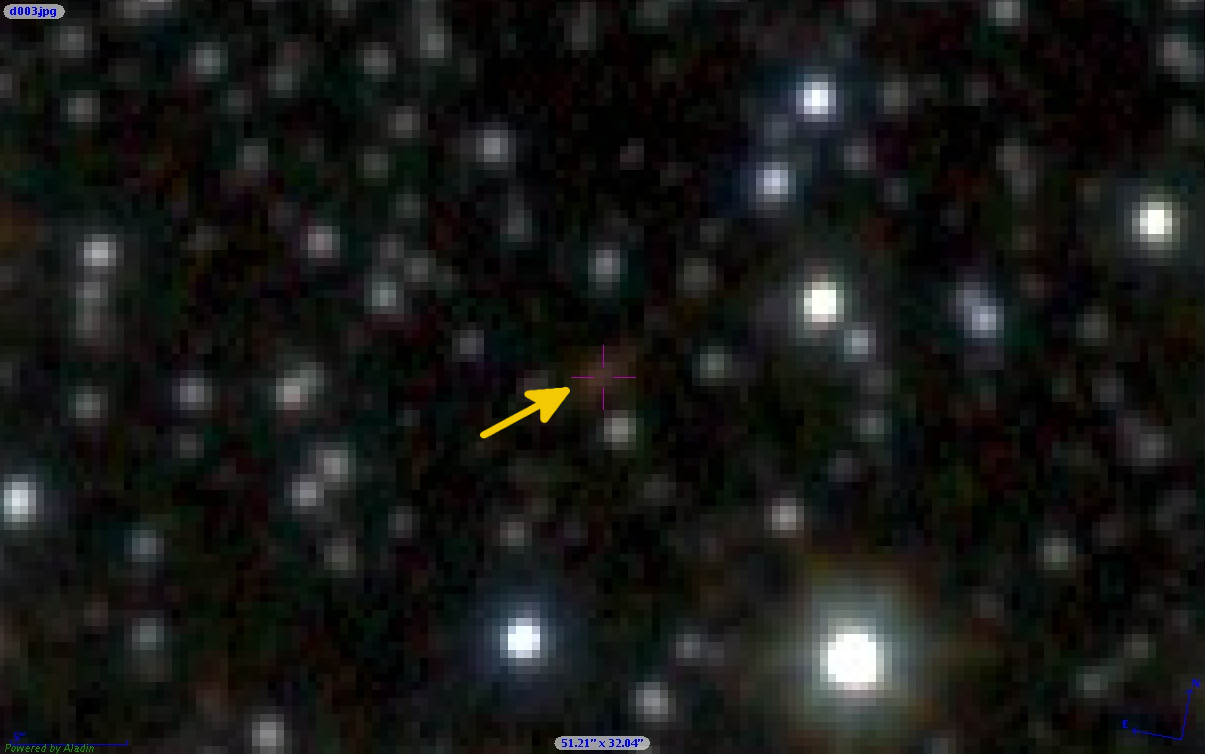} &
\includegraphics[bb=0 0 868 543,width=3.0cm]{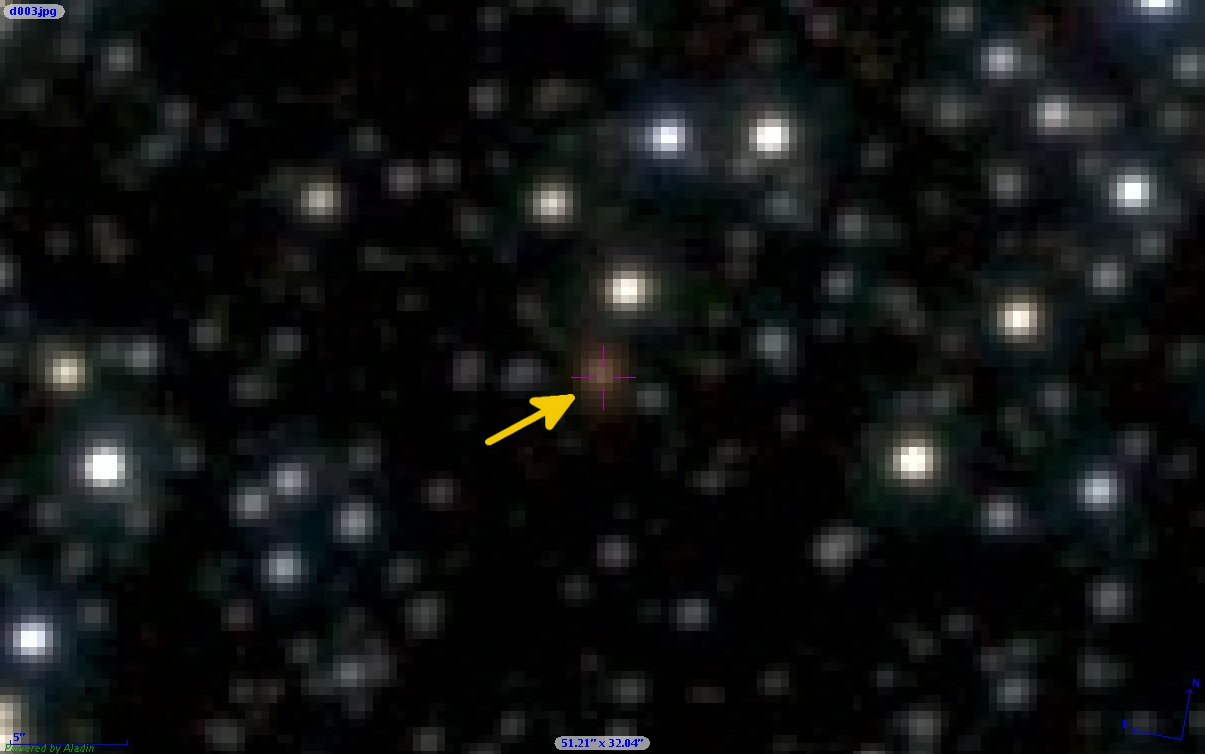} &
\includegraphics[bb=0 0 868 543,width=3.0cm]{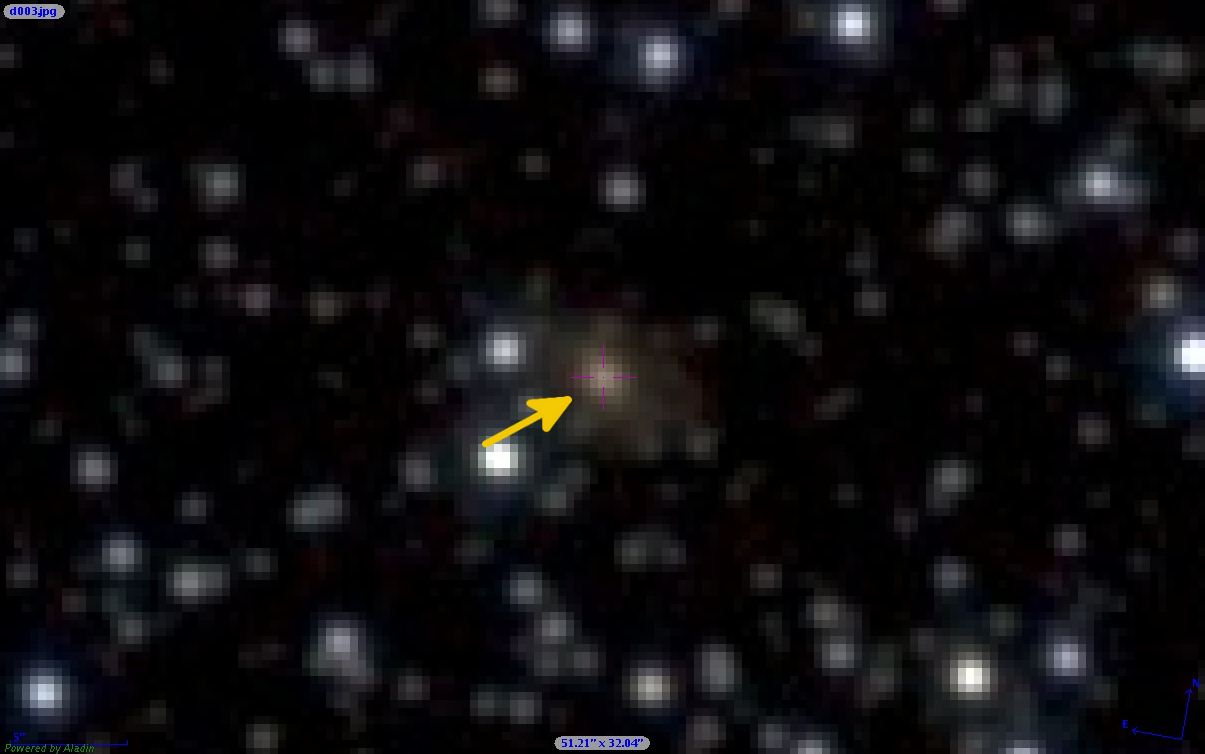} \\ 
205 & 208 & 213 & 226 \\ \\ \\
\includegraphics[bb=0 0 868 543,width=3.0cm]{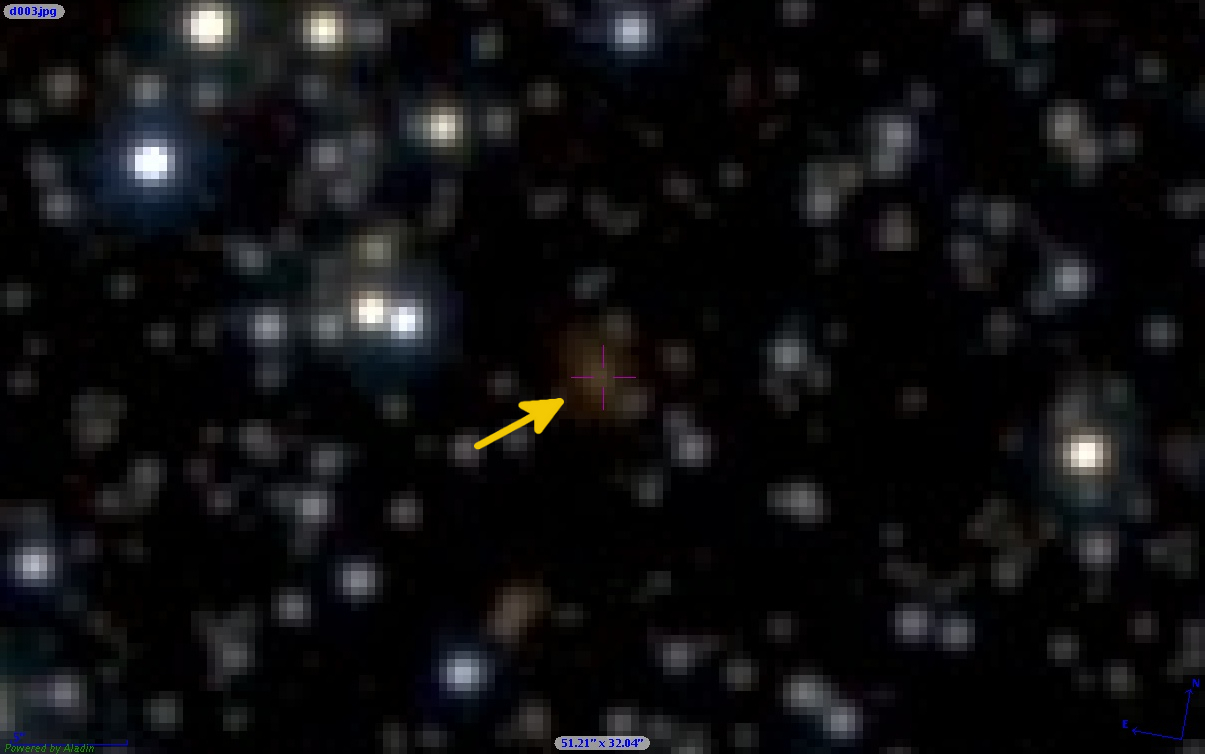} &
\includegraphics[bb=0 0 868 543,width=3.0cm]{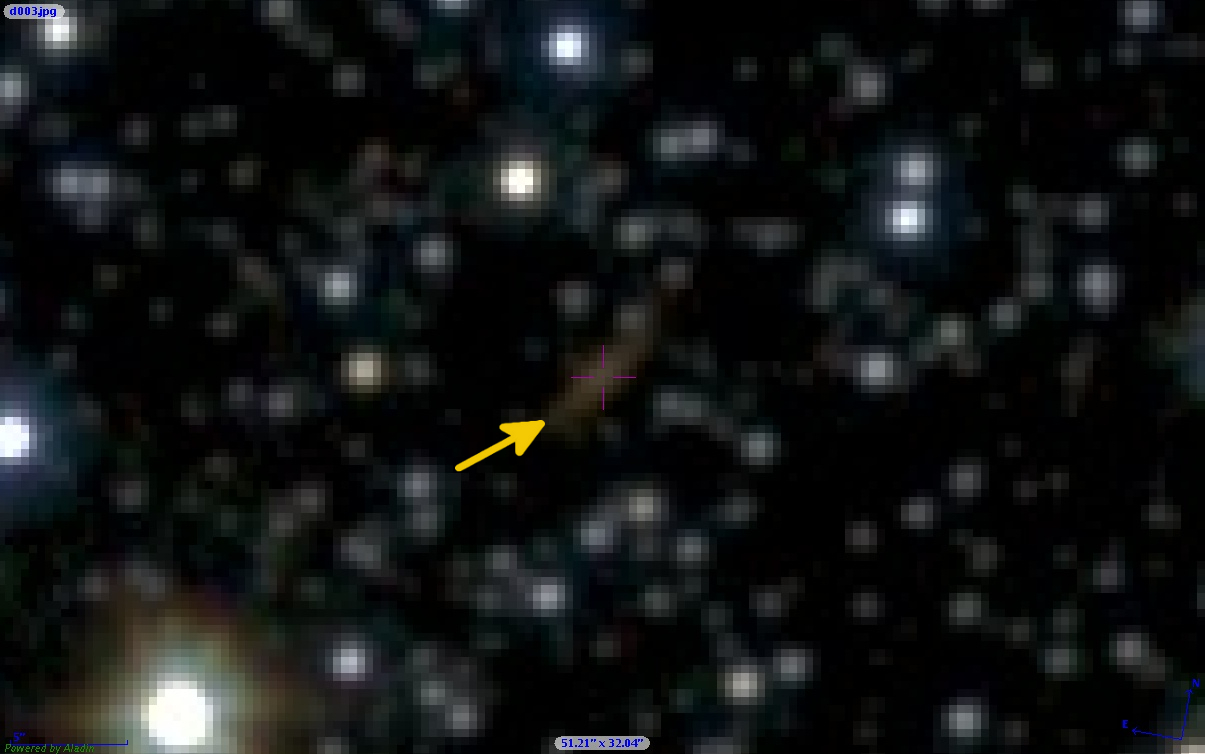} &
\includegraphics[bb=0 0 868 543,width=3.0cm]{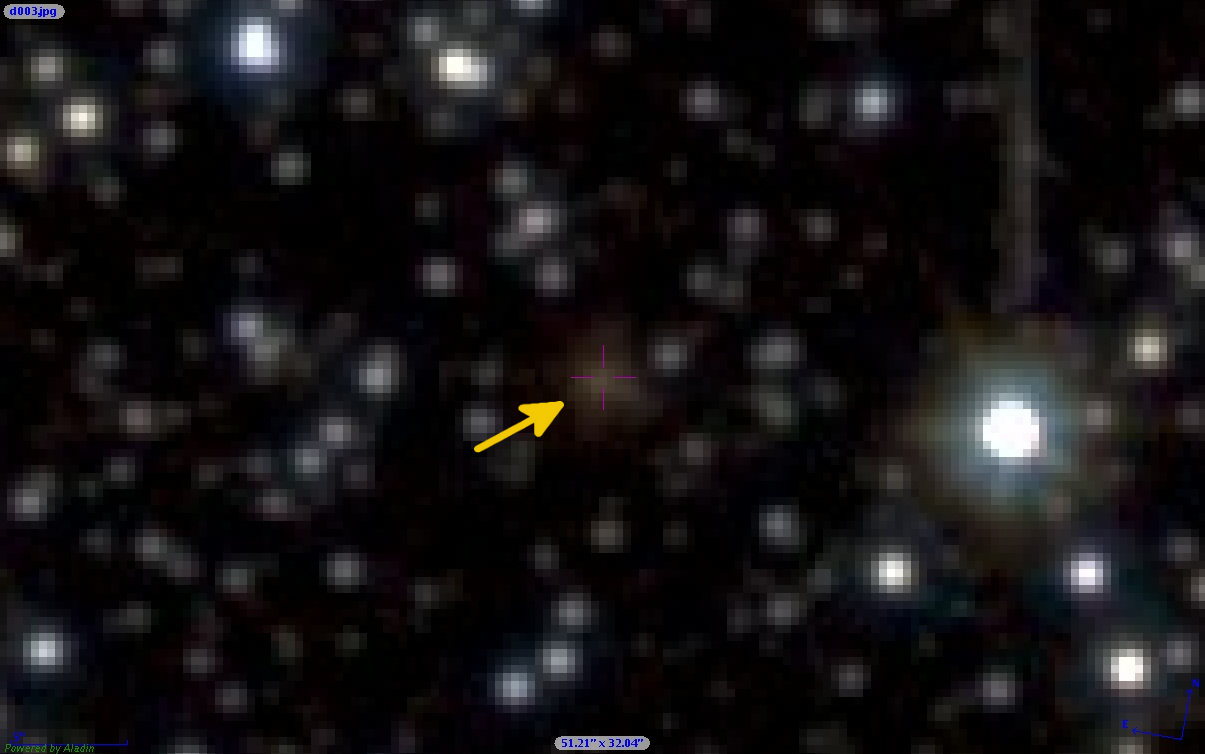} &
\includegraphics[bb=0 0 868 543,width=3.0cm]{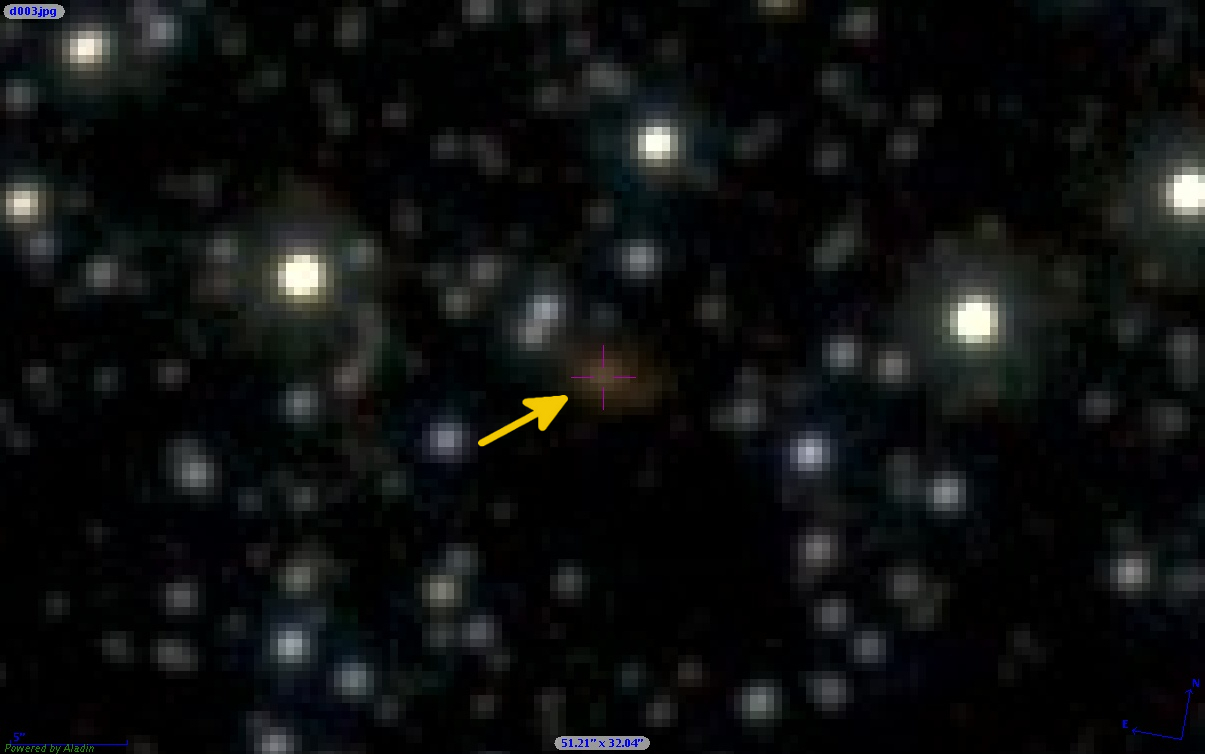} \\ 
229 & 244 & 249 & 257 \\ \\ \\ 
\includegraphics[bb=0 0 868 543,width=3.0cm]{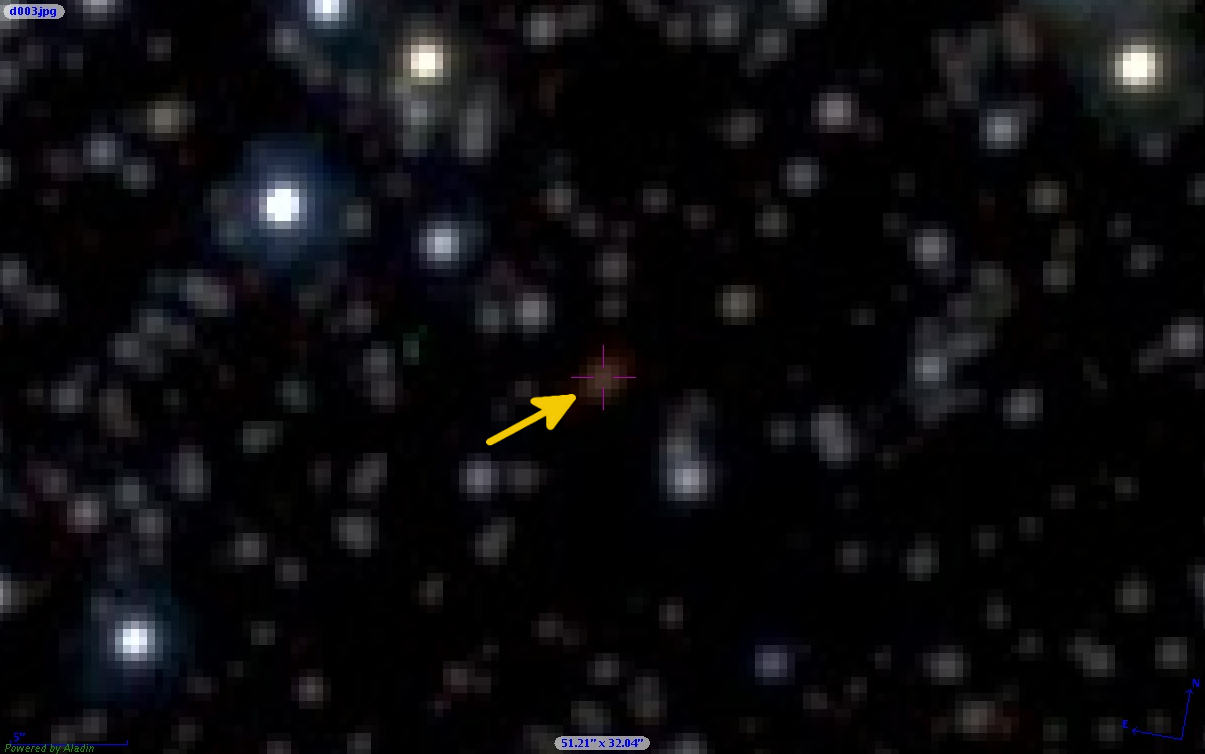} &
\includegraphics[bb=0 0 868 543,width=3.0cm]{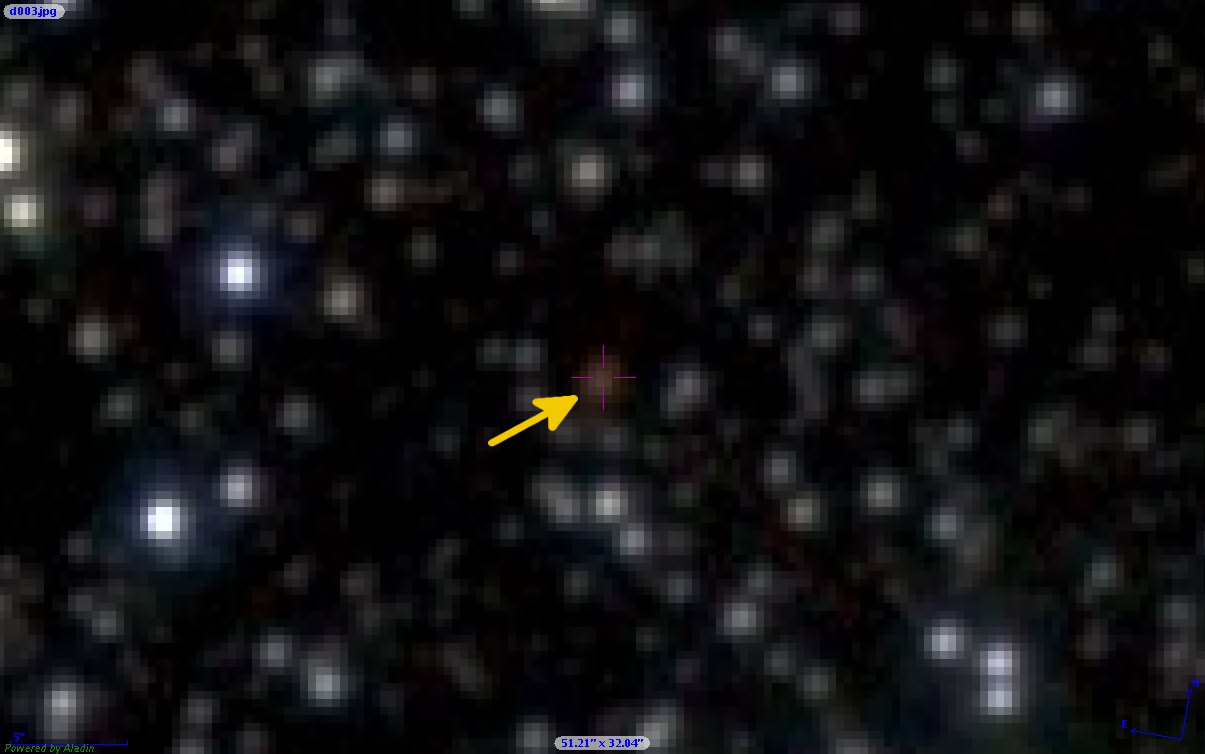} &
\includegraphics[bb=0 0 868 543,width=3.0cm]{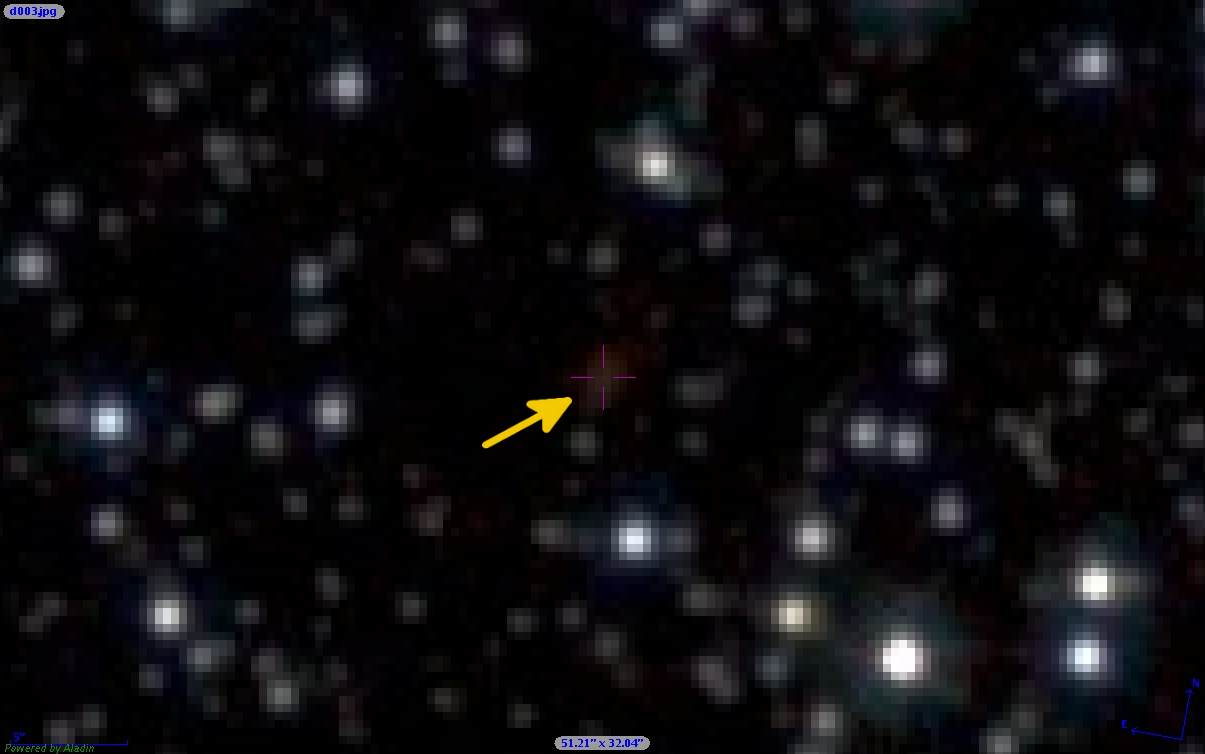} &
\includegraphics[bb=0 0 868 543,width=3.0cm]{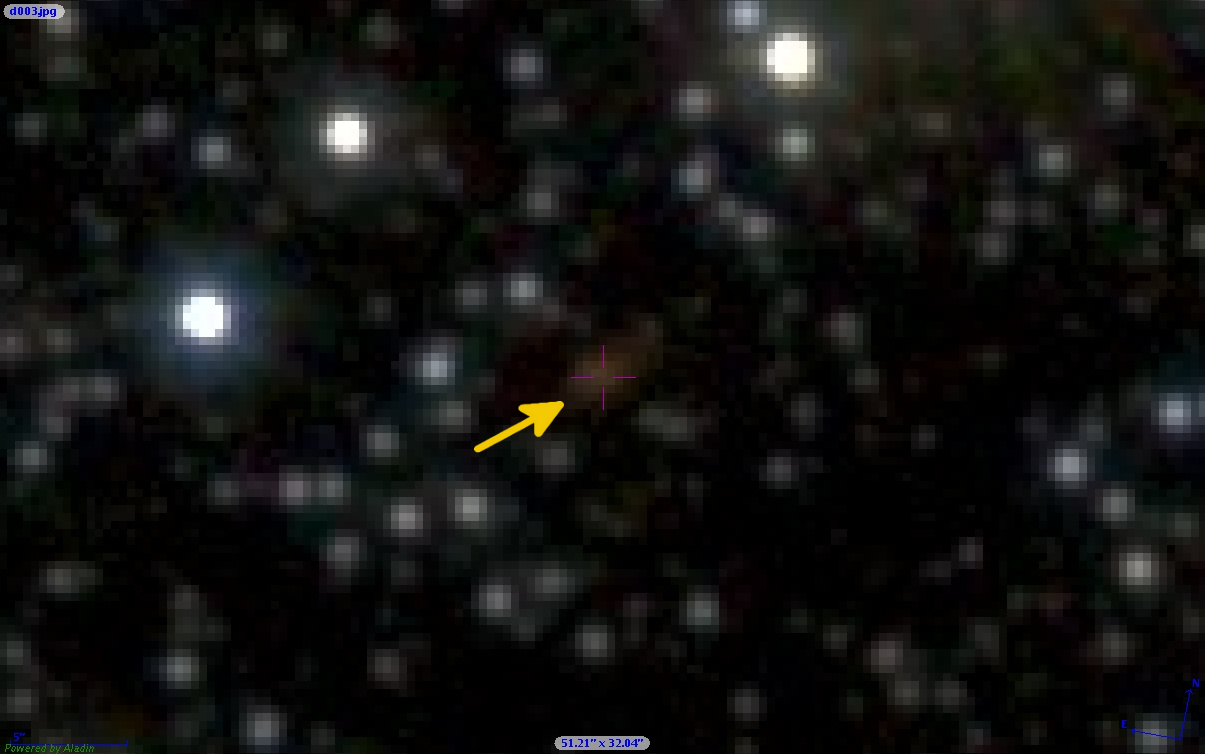} \\
\end{tabular}
\end{minipage}
\caption{False-color image from J, H \& K$_s$ passbands in
equatorial coordinates J2000 (RA,DEC), for 20 objects classified
by us as galaxy candidates that are located in the center of each
figure (yellow arrow). The numbers follow the numbering as
provided our catalog. The size of images are 51.2" and 32.1" in X
and Y axis, respectively.}
\end{figure*}

\clearpage
\newpage

\clearpage

\begin{figure*}
\begin{minipage}{1.0\textwidth}
\includegraphics[angle=90,width=16.0cm]{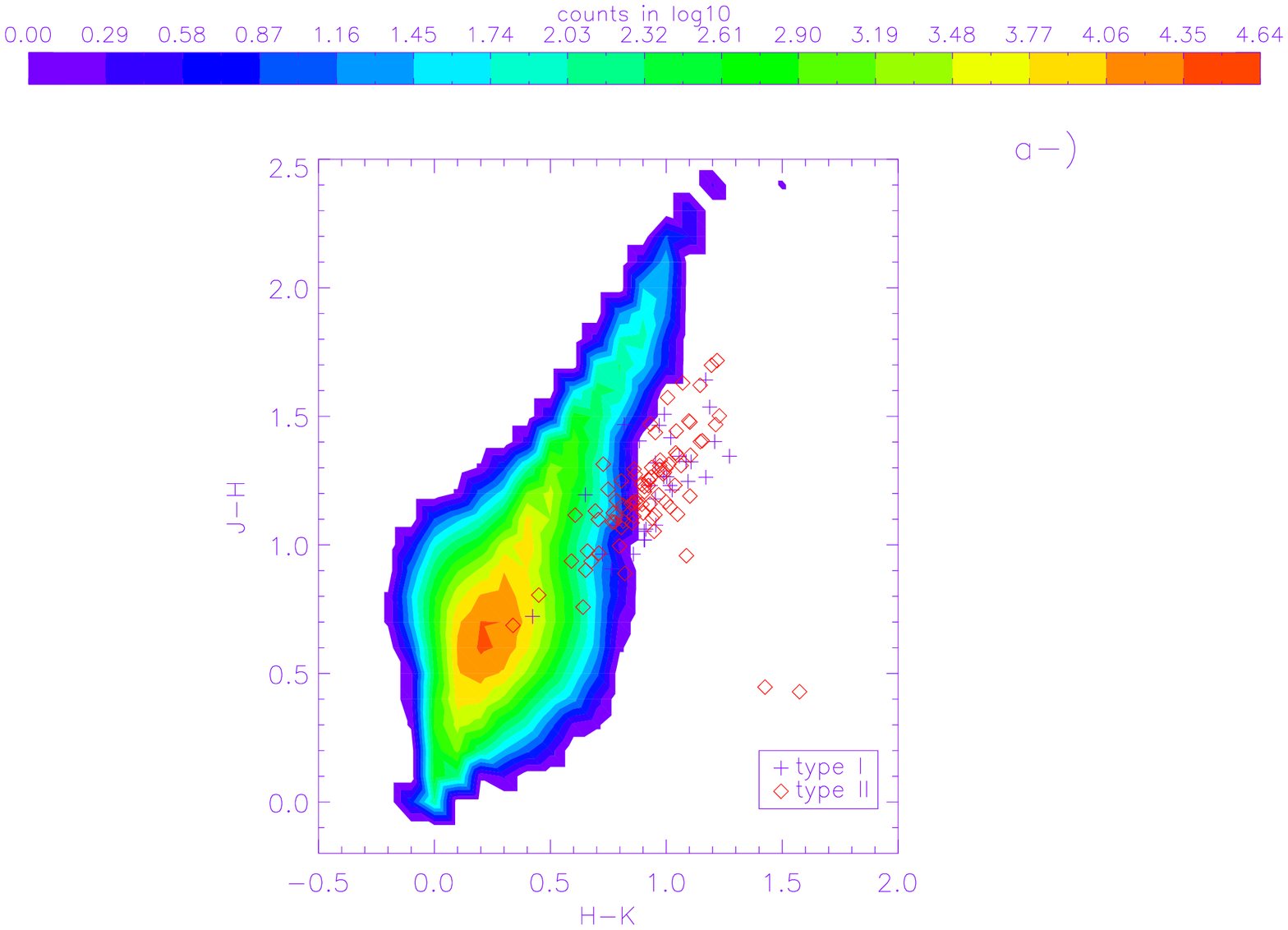}\\
\includegraphics[angle=90,width=16.0cm]{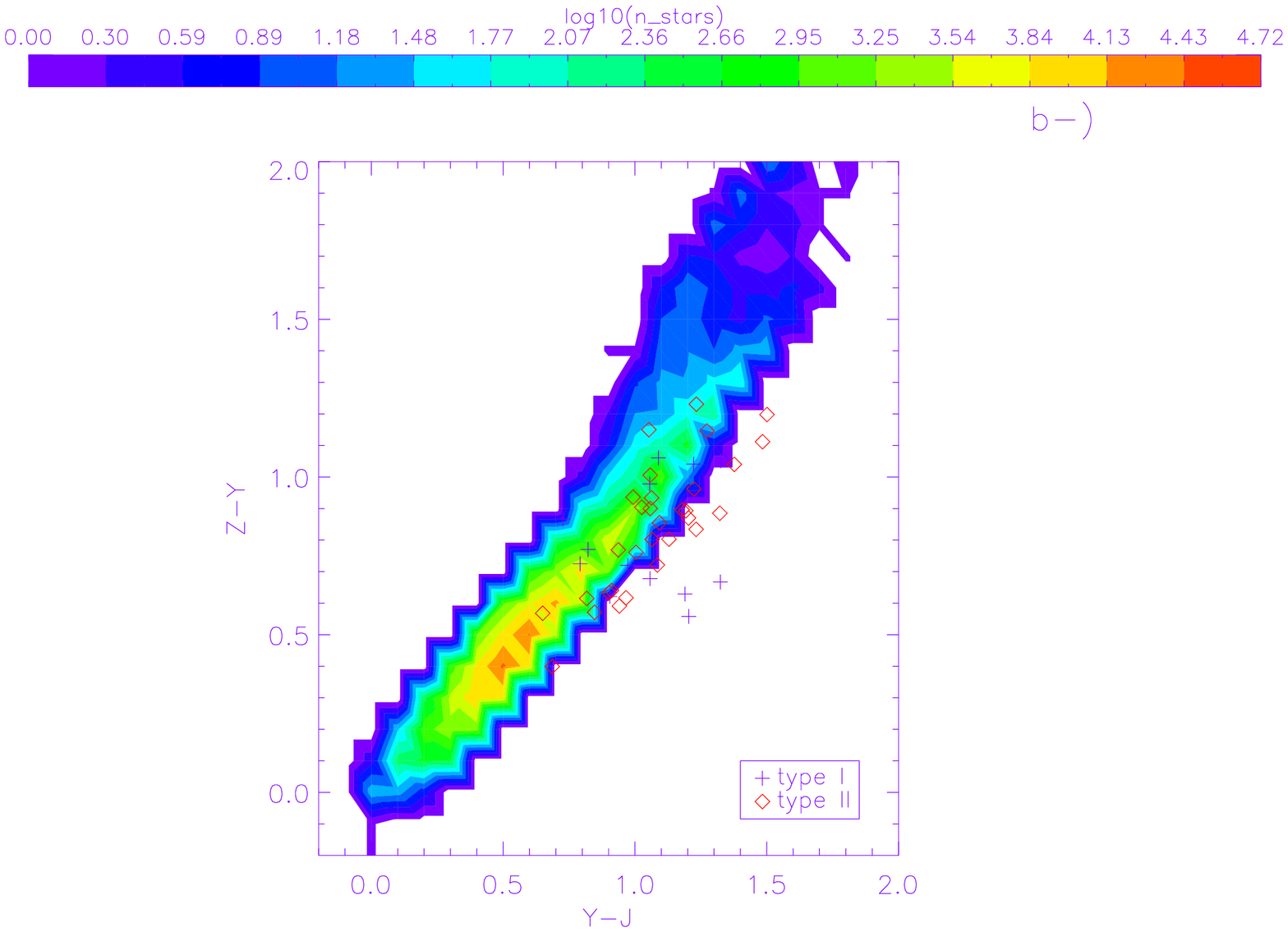}\\
\end{minipage}
\caption[]{}
\end{figure*}
\begin{figure*}
\ContinuedFloat
\begin{minipage}{1.0\textwidth}
\includegraphics[angle=90,width=\textwidth]{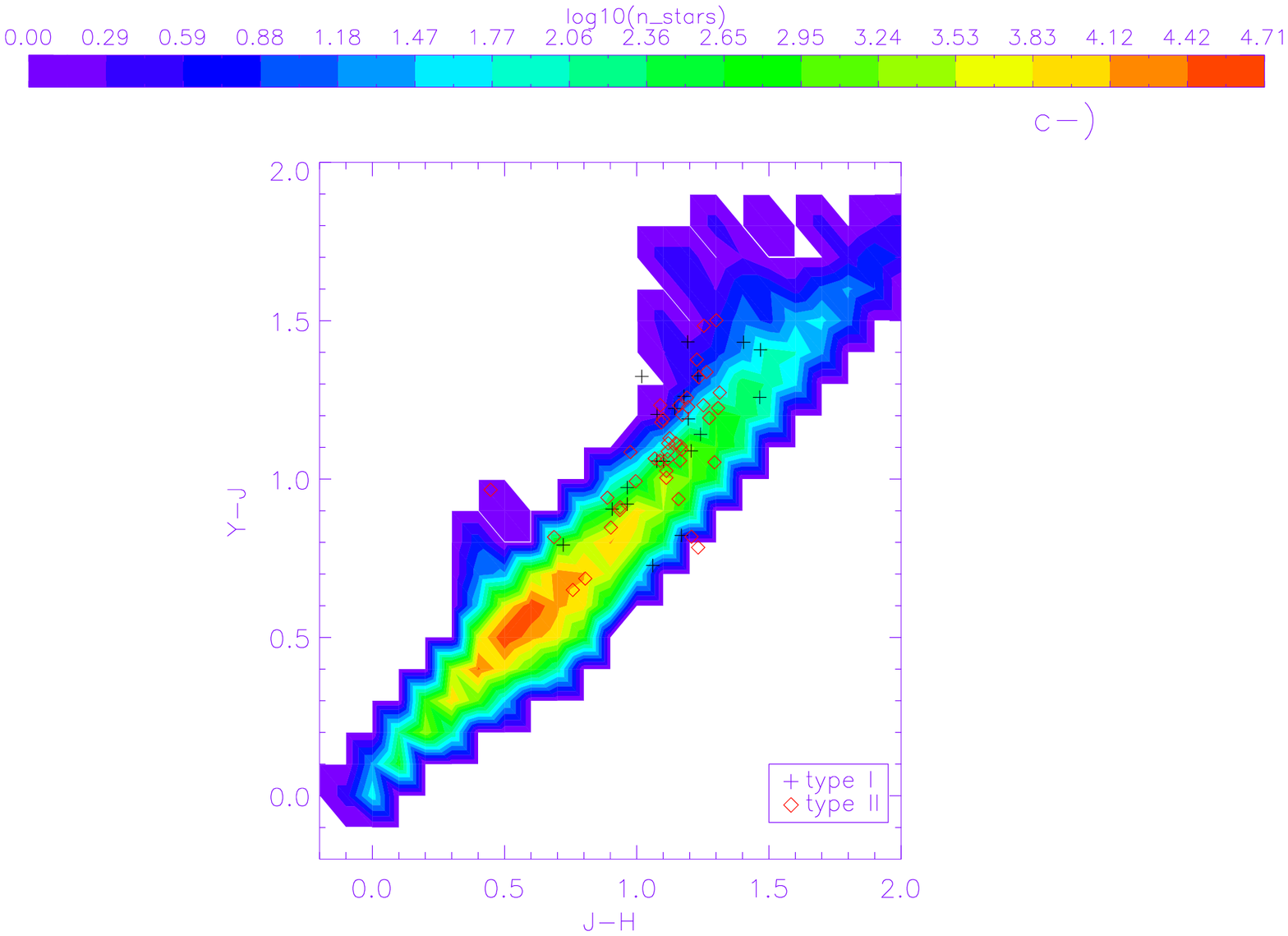}
\end{minipage}
\caption{Color-color density maps obtained from star counts
simulations (BGM and TRI-LEGAL models) for entire d003 tile. The
color bars are star counts for boxes with size equal to 0.1 x 0.1
square in color. The symbols represent source Type I (crosses) and
source Type II (red diamonds).}
\end{figure*}

\clearpage
\newpage

\begin{figure*}
\begin{minipage}{2.0\textwidth}
\begin{tabular}{cccc}
5 & 6 & 10 & 17 \\ \\ \\ \\
\includegraphics[bb=0 0 65 65,width=3.0cm]{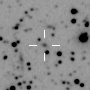} &
\includegraphics[bb=0 0 65 65,width=3.0cm]{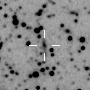} &
\includegraphics[bb=0 0 65 65,width=3.0cm]{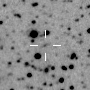} &
\includegraphics[bb=0 0 65 65,width=3.0cm]{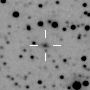} \\ 
24 & 25 & 33 & 41 \\ \\ \\ \\
\includegraphics[bb=0 0 65 65,width=3.0cm]{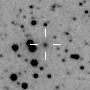} &
\includegraphics[bb=0 0 65 65,width=3.0cm]{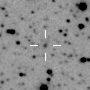} &
\includegraphics[bb=0 0 65 65,width=3.0cm]{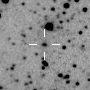} &
\includegraphics[bb=0 0 65 65,width=3.0cm]{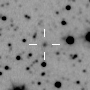} \\
42 & 44 & 46 & 48 \\ \\ \\ \\
\includegraphics[bb=0 0 65 65,width=3.0cm]{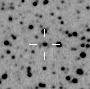} &
\includegraphics[bb=0 0 65 65,width=3.0cm]{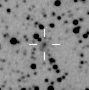} &
\includegraphics[bb=0 0 65 65,width=3.0cm]{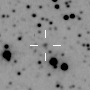} &
\includegraphics[bb=0 0 65 65,width=3.0cm]{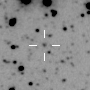} \\
49 & 57 & 58 & 60 \\ \\ \\ \\
\includegraphics[bb=0 0 65 65,width=3.0cm]{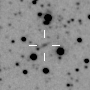} &
\includegraphics[bb=0 0 65 65,width=3.0cm]{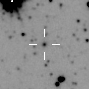} &
\includegraphics[bb=0 0 65 65,width=3.0cm]{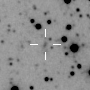} &
\includegraphics[bb=0 0 65 65,width=3.0cm]{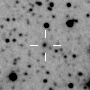} \\
\end{tabular}
\end{minipage}
\caption[]{}
\end{figure*}
\begin{figure*}
\ContinuedFloat
\begin{minipage}{2.0\textwidth}
\begin{tabular}{cccc}
65 & 69 & 70 & 72 \\ \\ \\ \\
\includegraphics[bb=0 0 65 65,width=3.0cm]{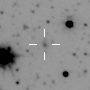} &
\includegraphics[bb=0 0 65 65,width=3.0cm]{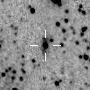} &
\includegraphics[bb=0 0 65 65,width=3.0cm]{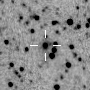} &
\includegraphics[bb=0 0 65 65,width=3.0cm]{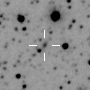} \\ 
74 & 76 & 77 & 79 \\ \\ \\ \\
\includegraphics[bb=0 0 65 65,width=3.0cm]{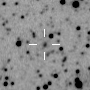} &
\includegraphics[bb=0 0 65 65,width=3.0cm]{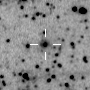} &
\includegraphics[bb=0 0 65 65,width=3.0cm]{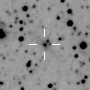} &
\includegraphics[bb=0 0 65 65,width=3.0cm]{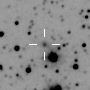} \\
84 & 86 & 91 & 92 \\ \\ \\ \\
\includegraphics[bb=0 0 65 65,width=3.0cm]{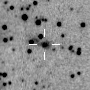} &
\includegraphics[bb=0 0 65 65,width=3.0cm]{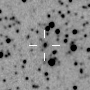} &
\includegraphics[bb=0 0 65 65,width=3.0cm]{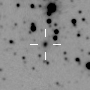} &
\includegraphics[bb=0 0 65 65,width=3.0cm]{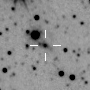} \\ 
93 & 94 & 107 & 108 \\ \\ \\ \\
\includegraphics[bb=0 0 65 65,width=3.0cm]{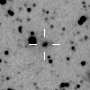} &
\includegraphics[bb=0 0 65 65,width=3.0cm]{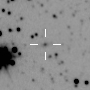} &
\includegraphics[bb=0 0 65 65,width=3.0cm]{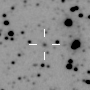} &
\includegraphics[bb=0 0 65 65,width=3.0cm]{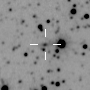} \\ 
\end{tabular}
\end{minipage}
\caption[]{}
\end{figure*}
\begin{figure*}
\ContinuedFloat
\begin{minipage}{2.0\textwidth}
\begin{tabular}{cccc}
115 & 128 & 133 & 135 \\ \\ \\ \\
\includegraphics[bb=0 0 65 65,width=3.0cm]{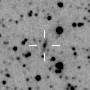} &
\includegraphics[bb=0 0 65 65,width=3.0cm]{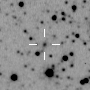} &
\includegraphics[bb=0 0 65 65,width=3.0cm]{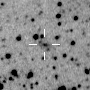} &
\includegraphics[bb=0 0 65 65,width=3.0cm]{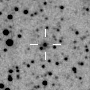} \\
146 & 147 & 150 & 151 \\ \\ \\ \\
\includegraphics[bb=0 0 65 65,width=3.0cm]{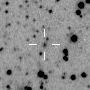} &
\includegraphics[bb=0 0 65 65,width=3.0cm]{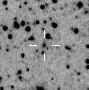} &
\includegraphics[bb=0 0 65 65,width=3.0cm]{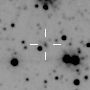} &
\includegraphics[bb=0 0 65 65,width=3.0cm]{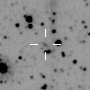} \\
161 & 162 & 165 & 167 \\ \\ \\ \\
\includegraphics[bb=0 0 65 65,width=3.0cm]{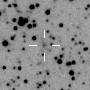} &
\includegraphics[bb=0 0 65 65,width=3.0cm]{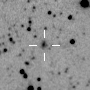} &
\includegraphics[bb=0 0 65 65,width=3.0cm]{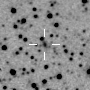} &
\includegraphics[bb=0 0 65 65,width=3.0cm]{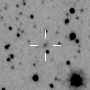} \\
168 & 173 & 177 & 178 \\ \\ \\ \\
\includegraphics[bb=0 0 65 65,width=3.0cm]{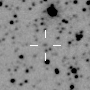} &
\includegraphics[bb=0 0 65 65,width=3.0cm]{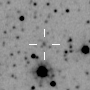} &
\includegraphics[bb=0 0 65 65,width=3.0cm]{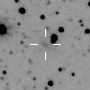} &
\includegraphics[bb=0 0 65 65,width=3.0cm]{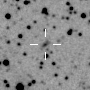} \\
\end{tabular}
\end{minipage}
\caption[]{}
\end{figure*}
\begin{figure*}
\ContinuedFloat
\begin{minipage}{2.0\textwidth}
\begin{tabular}{cccc}
182 & 187 & 189 & 191 \\ \\ \\ \\
\includegraphics[bb=0 0 65 65,width=3.0cm]{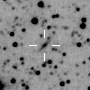} &
\includegraphics[bb=0 0 65 65,width=3.0cm]{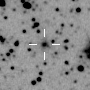} &
\includegraphics[bb=0 0 65 65,width=3.0cm]{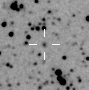} &
\includegraphics[bb=0 0 65 65,width=3.0cm]{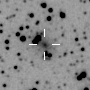} \\
196 & 197 & 199 & 201 \\ \\ \\ \\
\includegraphics[bb=0 0 65 65,width=3.0cm]{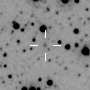} &
\includegraphics[bb=0 0 65 65,width=3.0cm]{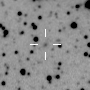} &
\includegraphics[bb=0 0 65 65,width=3.0cm]{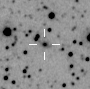} &
\includegraphics[bb=0 0 65 65,width=3.0cm]{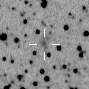} \\
202 & 208 & 211 & 216 \\ \\ \\ \\
\includegraphics[bb=0 0 65 65,width=3.0cm]{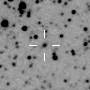} &
\includegraphics[bb=0 0 65 65,width=3.0cm]{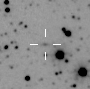} &
\includegraphics[bb=0 0 65 65,width=3.0cm]{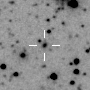} &
\includegraphics[bb=0 0 65 65,width=3.0cm]{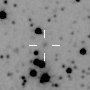} \\
217 & 218 & 220 & 226 \\ \\ \\ \\
\includegraphics[bb=0 0 65 65,width=3.0cm]{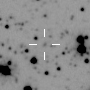} &
\includegraphics[bb=0 0 65 65,width=3.0cm]{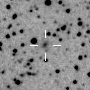} &
\includegraphics[bb=0 0 65 65,width=3.0cm]{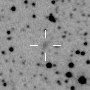} &
\includegraphics[bb=0 0 65 65,width=3.0cm]{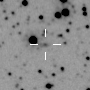} \\
\end{tabular}
\end{minipage}
\caption[]{}
\end{figure*}
\begin{figure*}
\ContinuedFloat
\begin{minipage}{2.0\textwidth}
\begin{tabular}{cccc}
227 & 228 & 239 & 240 \\ \\ \\ \\
\includegraphics[bb=0 0 65 65,width=3.0cm]{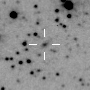} &
\includegraphics[bb=0 0 65 65,width=3.0cm]{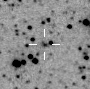} &
\includegraphics[bb=0 0 65 65,width=3.0cm]{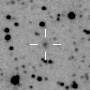} &
\includegraphics[bb=0 0 65 65,width=3.0cm]{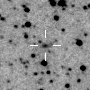} \\
245 & 259 & 260 & 262 \\ \\ \\ \\
\includegraphics[bb=0 0 65 65,width=3.0cm]{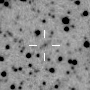} &
\includegraphics[bb=0 0 65 65,width=3.0cm]{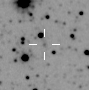} &
\includegraphics[bb=0 0 65 65,width=3.0cm]{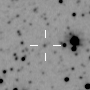} &
\includegraphics[bb=0 0 65 65,width=3.0cm]{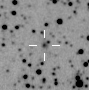} \\
\end{tabular}
\end{minipage}
\caption{Images of objects (white crosses in the image center)
classified by us as possible galaxies in the K$_s$ filter with our
galaxy number presented for each galaxy. The image sizes are 30"
arcsec in Equatorial coordinates J2000 (RA,DEC).}
\end{figure*}

\clearpage

\begin{figure*}
\begin{minipage}{2.0\textwidth}
\begin{tabular}{ccccc}
Z & Y & J & H & Ks \\ \\ \\ \\
\includegraphics[bb=0 0 65 65,width=2.5cm]{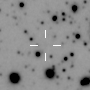}&
\includegraphics[bb=0 0 65 65,width=2.5cm]{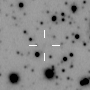}&
\includegraphics[bb=0 0 65 65,width=2.5cm]{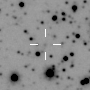}&
\includegraphics[bb=0 0 65 65,width=2.5cm]{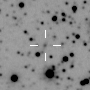}&
\includegraphics[bb=0 0 65 65,width=2.5cm]{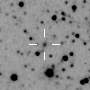} \\ 
128 \\ \\ \\ \\
\includegraphics[bb=0 0 65 65,width=2.5cm]{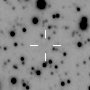}&
\includegraphics[bb=0 0 65 65,width=2.5cm]{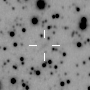}&
\includegraphics[bb=0 0 65 65,width=2.5cm]{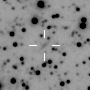}&
\includegraphics[bb=0 0 65 65,width=2.5cm]{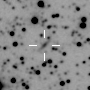}&
\includegraphics[bb=0 0 65 65,width=2.5cm]{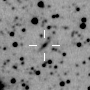} \\ 
182 \\ \\ \\ \\
\includegraphics[bb=0 0 65 65,width=2.5cm]{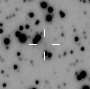}&
\includegraphics[bb=0 0 65 65,width=2.5cm]{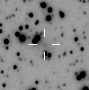}&
\includegraphics[bb=0 0 65 65,width=2.5cm]{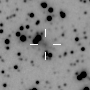}&
\includegraphics[bb=0 0 65 65,width=2.5cm]{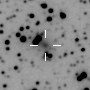}&
\includegraphics[bb=0 0 65 65,width=2.5cm]{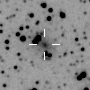} \\ 
191 \\ \\ \\ \\
\includegraphics[bb=0 0 65 65,width=2.5cm]{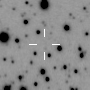}&
\includegraphics[bb=0 0 65 65,width=2.5cm]{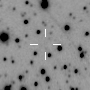}&
\includegraphics[bb=0 0 65 65,width=2.5cm]{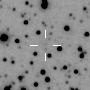}&
\includegraphics[bb=0 0 65 65,width=2.5cm]{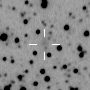}&
\includegraphics[bb=0 0 65 65,width=2.5cm]{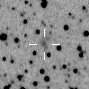} \\ 
201 \\ \\ \\ \\
\includegraphics[bb=0 0 65 65,width=2.5cm]{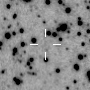} &
\includegraphics[bb=0 0 65 65,width=2.5cm]{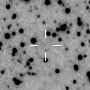} &
\includegraphics[bb=0 0 65 65,width=2.5cm]{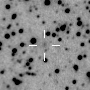} &
\includegraphics[bb=0 0 65 65,width=2.5cm]{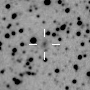} &
\includegraphics[bb=0 0 65 65,width=2.5cm]{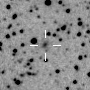} \\ 
218 \\ \\ \\
\end{tabular}
\end{minipage}
\caption{Multi-band image in ZYJH$K_s$ for objects classified as
source Type I objects. The image sizes are 30" arcsec in
Equatorial coordinates J2000 (RA,DEC).}
\end{figure*}

\clearpage

\begin{figure*}[h]
\centering
\resizebox{13.0cm}{!}{\includegraphics[angle=90,scale=1.0]{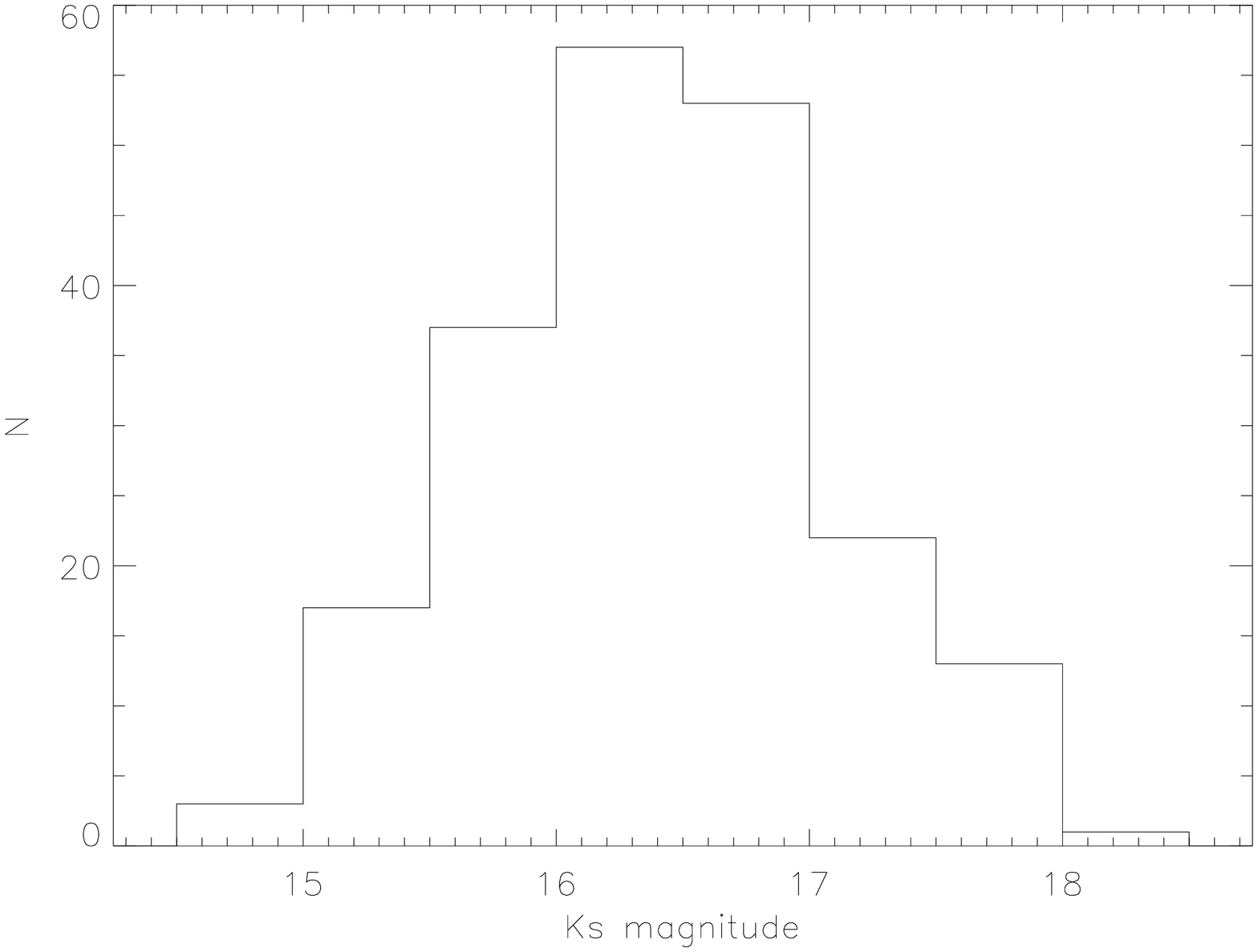}}
\caption{K$_s$ counts for objects source Type I and II.}
\end{figure*}

\clearpage

\begin{figure*}[h]
\centering
\resizebox{16.0cm}{!}{\includegraphics[angle=90,scale=1.0]{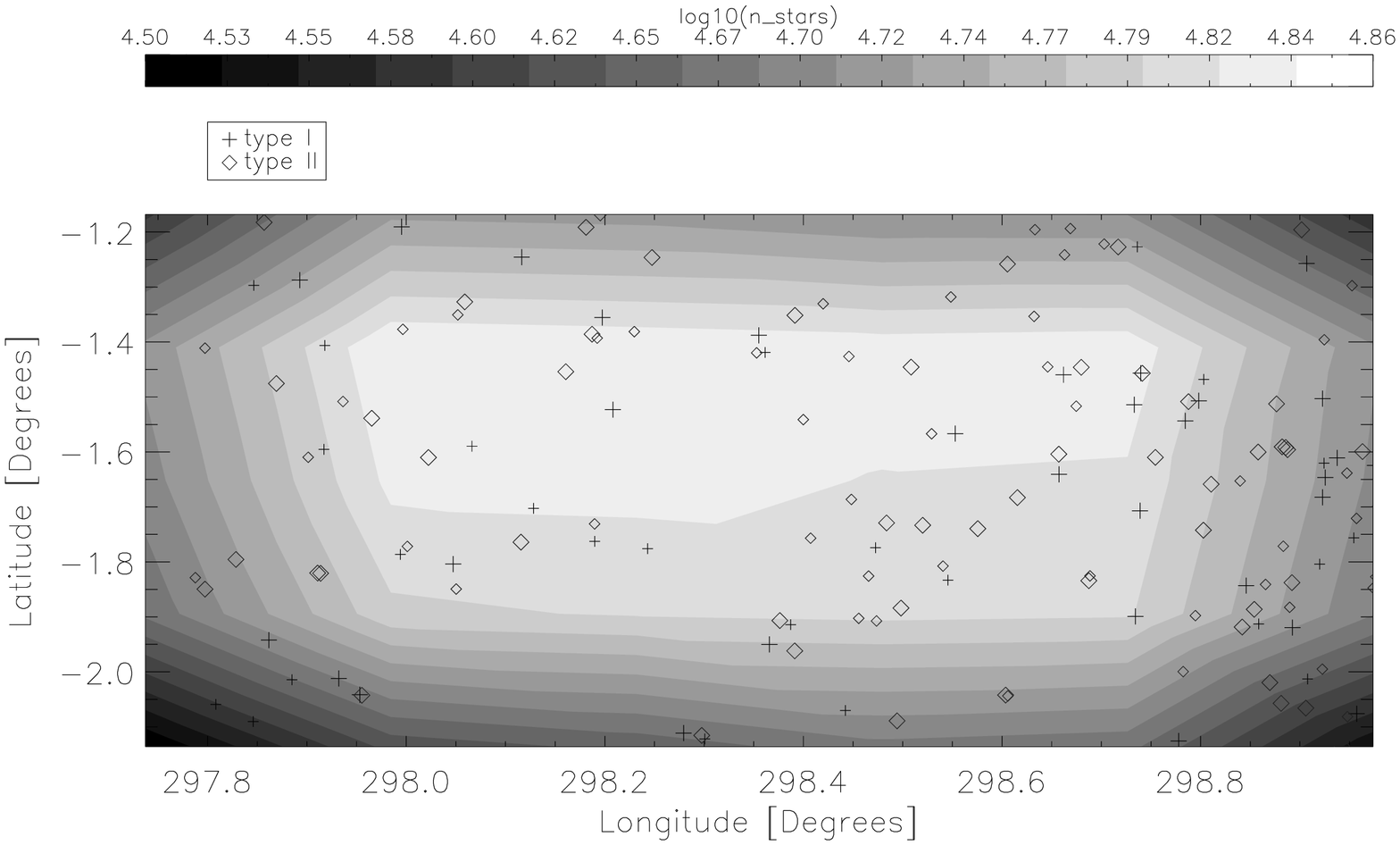}}
\caption{VVV Density counts in galactic coordinates for sources
classified as stellar in the d003 tile for K$_s$ brighter than or
equal to 15.0 mag. Symbols are source Type I (crosses) and II
(diamonds). Smaller/bigger symbols represent different magnitudes
ranges K$_s \leq 16.5$ and K$_s > 16.5$, respectively.}
\end{figure*}

\begin{figure*}[h]
\centering
\resizebox{15.0cm}{!}{\includegraphics[angle=90,scale=1.0]{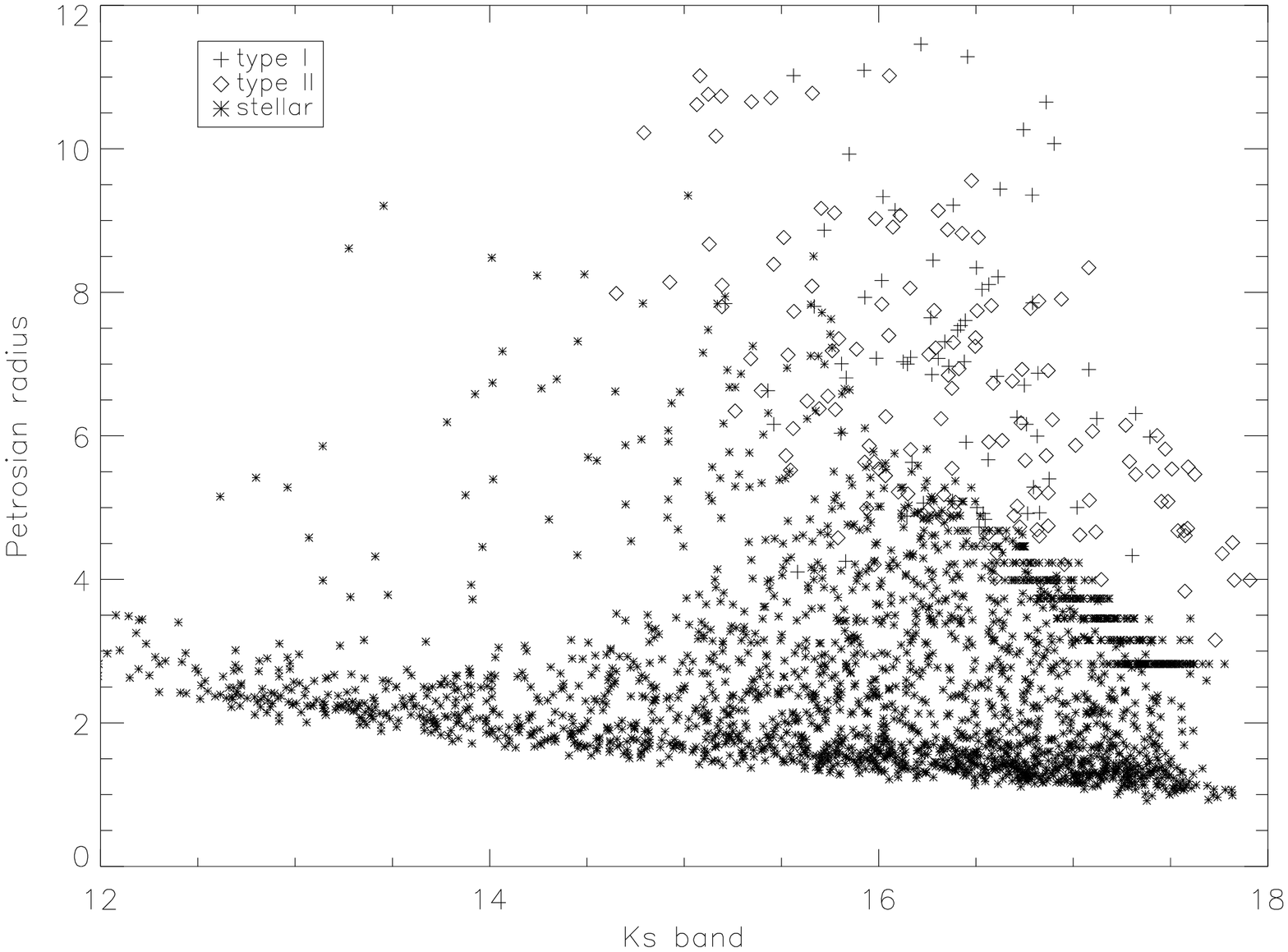}}
\caption{Size distribution of the galaxy candidates (Petrosian
radius versus magnitude): source Type I (crosses), source Type II
(diamonds) and stars (asterisks.)}
\end{figure*}

\clearpage

\begin{figure*}[h]
\centering \resizebox{13.0cm}{!}{\includegraphics{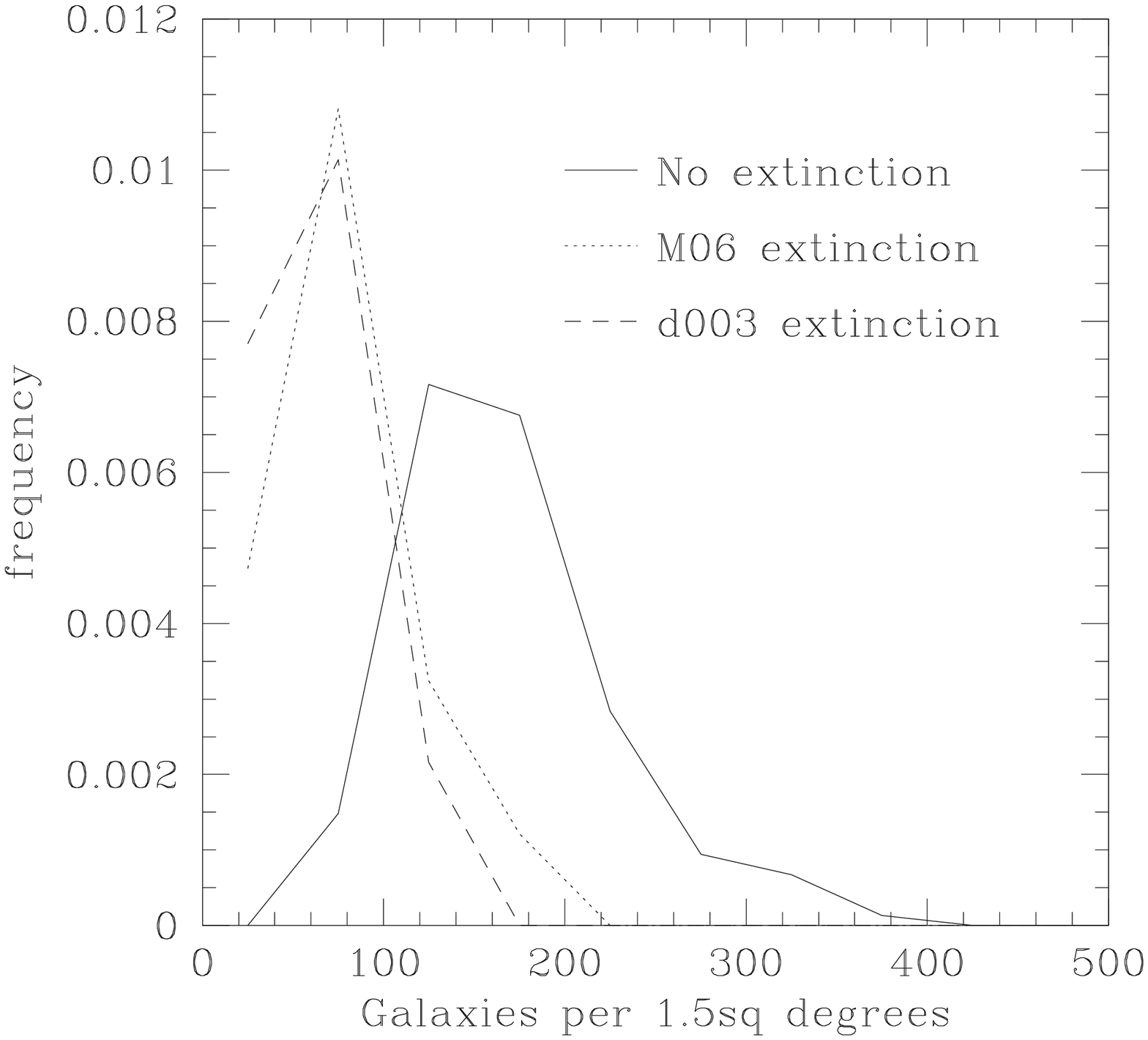}}
\caption{Histogram of number of galaxies per disc VVV tile of
$1.5$ sq. degrees from the Mock catalogues (solid lines for the
catalogue with no Galactic extinction, dotted when the extinction
is taken into account), dashed lines is the average extinction for
d003.}
\end{figure*}

\bigskip
\begin{table*}
\centering \caption{Objects observed by 2MASS as extended sources
for three different regions of the VVV survey; (\textit{vc}) means
verification of source equal to -2, -1, 1 and 2, which corresponds
to unknown, no verification, galaxy, non$-$extended (e.g., star,
double, triple, artifact), respectively.}
\begin{flushleft}
\begin{tabular}{ccccc}
\hline region &  $vc=-2$  &  $-1$  & $1$ & $2$ \\
\hline \noalign{\smallskip}
VVV disk &  22 & 0  & 155 & 0 \\
VVV bulge&  46 & 11 & 187 & 6 \\
$295^{o} \leq \ell \leq 300^{o}$ $|b| \le 4^{o}$& 14 & 4 & 65 & 1\\
\noalign{\smallskip} \hline
\end{tabular}
\end{flushleft}
\end{table*}
\bigskip

\begin{landscape}
\tiny
\begin{center}
\onecolumn
\begin{longtable}{lccccccccccc}
\caption{{\scriptsize \sl Galaxy candidates properties. Class is our object
classification: 1 (source Type I) and 2 (source Type II).}} \label{gallist} \\

\hline \hline \\[-2ex]
  \multicolumn{1}{c}{n$^{\rm{o}}$} &
  \multicolumn{1}{c}{Gal. Longitude} &
  \multicolumn{1}{c}{Gal. Latitude} &
  \multicolumn{1}{c}{Z} &
  \multicolumn{1}{c}{Y} &
  \multicolumn{1}{c}{J} &
  \multicolumn{1}{c}{H} &
  \multicolumn{1}{c}{K$_s$} &
  \multicolumn{1}{c}{$r_{\rm{Petrosian}}$} &
  \multicolumn{1}{c}{ellipticity} &
  \multicolumn{1}{c}{PA} &
  \multicolumn{1}{c}{class} \\[0.5ex] \hline \\[-1.8ex]
\endfirsthead

\multicolumn{12}{c}{{\tablename} \thetable{} -- Continued} \\[0.5ex]
\hline \hline \\[-2ex]
  \multicolumn{1}{c}{n$^{\rm{o}}$} &
  \multicolumn{1}{c}{Gal. Longitude} &
  \multicolumn{1}{c}{Gal. Latitude} &
  \multicolumn{1}{c}{Z} &
  \multicolumn{1}{c}{Y} &
  \multicolumn{1}{c}{J} &
  \multicolumn{1}{c}{H} &
  \multicolumn{1}{c}{K$_s$} &
  \multicolumn{1}{c}{$r_{\rm{Petrosian}}$} &
  \multicolumn{1}{c}{ellipticity} &
  \multicolumn{1}{c}{PA} &
  \multicolumn{1}{c}{class} \\[0.5ex] \hline \\[-1.8ex]
\endhead

\multicolumn{12}{l}{{Continued on Next Page\ldots}} \\
\endfoot

\\[-1.8ex] \hline \hline
\endlastfoot

  1 &   299.054643 &  -1.216376 & ------ $\pm$ ------& ------$\pm$ ------&
------$\pm$ ------& 16.689$\pm$  0.033& 16.354$\pm$ 0.049&   8.87&  0.183&
117.5& 2\\
  2 &   299.052010 &  -1.248499 & 19.330$\pm$  0.074& 18.715$\pm$  0.071& 17.898$\pm$  0.054& 17.211$\pm$  0.054& 16.872$\pm$ 0.079&   5.20&  0.414&  41.6& 2\\
  3 &   299.047234 &  -1.294900 & 19.524$\pm$  0.089& 18.891$\pm$  0.083& 17.989$\pm$  0.059& 17.053$\pm$  0.046& 16.377$\pm$ 0.051&   5.54&  0.303&  86.9& 2\\
  4 &   299.046990 &  -1.579076 & 19.535$\pm$  0.088& 18.944$\pm$  0.086& 18.003$\pm$  0.059& 17.114$\pm$  0.048& 16.293$\pm$ 0.046&   7.22&  0.671& 109.6& 2\\
  5 &   299.047811 &  -1.580668 & ------$\pm$ ------& ------$\pm$ ------& 18.400$\pm$  0.084& 17.134$\pm$  0.049& 16.145$\pm$ 0.040&   4.88&  0.261& 107.9& 1\\
  6 &   299.050006 &  -1.582006 & ------$\pm$ ------& ------$\pm$ ------& 18.467$\pm$  0.090& 17.220$\pm$  0.053& 16.126$\pm$ 0.039&   7.03&  0.550& 169.3& 1\\
  9 &   298.984289 &  -1.470670 & 20.405$\pm$  0.193& ------$\pm$ ------& 18.104$\pm$  0.064& 16.937$\pm$  0.042& 15.938$\pm$ 0.033&   4.98&  0.163&  16.7& 2\\
 10 &   298.984422 &  -1.468328 & 20.056$\pm$  0.141& 19.331$\pm$  0.122& 18.539$\pm$  0.095& 17.817$\pm$  0.093& 17.394$\pm$ 0.125&   5.98&  0.483&  67.1& 1\\
 12 &   299.037877 &  -1.584041 & 19.946$\pm$  0.128& 18.906$\pm$  0.083& 17.529$\pm$  0.039& 16.302$\pm$  0.023& 15.396$\pm$ 0.020&   6.63&  0.108&  23.7& 2\\
 13 &   298.991020 &  -1.573740 & ------$\pm$ ------& ------$\pm$ ------& ------$\pm$ ------& 18.508$\pm$  0.174& 17.625$\pm$ 0.154&   5.46&  0.433&  16.1& 2\\
 14 &   299.000353 &  -1.675140 & 19.438$\pm$  0.080& 18.538$\pm$  0.060& 17.480$\pm$  0.037& 16.316$\pm$  0.024& 15.460$\pm$ 0.022&   8.39&  0.233&  82.9& 2\\
 15 &   298.985578 &  -1.679874 & ------$\pm$ ------& ------$\pm$ ------& 18.521$\pm$  0.094& 17.228$\pm$  0.054& 16.258$\pm$ 0.044&   7.13&  0.107&  79.1& 2\\
 16 &   298.999713 &  -1.680083 & 18.501$\pm$  0.035& 18.102$\pm$  0.040& 17.416$\pm$  0.035& 16.611$\pm$  0.031& 16.161$\pm$ 0.041&   8.05&  0.163&  74.4& 2\\
 17 &   298.978338 &  -1.698410 & ------$\pm$ ------& ------$\pm$ ------& 18.025$\pm$  0.060& 16.874$\pm$  0.039& ------$\pm$------&   4.95&  0.151& 999.9& 1\\
 19 &   299.031350 &  -1.782191 & ------$\pm$ ------& 19.297$\pm$  0.118& 18.184$\pm$  0.069& 17.065$\pm$  0.047& 16.016$\pm$ 0.036&   7.83&  0.374&  28.6& 2\\
 20 &   299.029333 &  -1.757842 & ------$\pm$ ------& ------$\pm$ ------& 18.383$\pm$  0.082& 17.259$\pm$  0.056& 16.358$\pm$ 0.048&   6.84&  0.277&  99.6& 2\\
 21 &   299.027119 &  -1.810838 & 18.314$\pm$  0.030& 17.743$\pm$  0.029& 16.896$\pm$  0.022& 15.994$\pm$  0.018& 15.342$\pm$ 0.019&   7.07&  0.246& 167.1& 2\\
 22 &   298.979133 &  -1.827110 & ------$\pm$ ------& ------$\pm$ ------& 17.224$\pm$  0.029& 16.256$\pm$  0.022& 15.547$\pm$ 0.023&   5.52&  0.121& 123.8& 2\\
 23 &   299.004038 &  -1.842581 & 19.639$\pm$  0.096& 18.743$\pm$  0.071& 17.564$\pm$  0.040& 16.471$\pm$  0.027& 15.657$\pm$ 0.026&   8.08&  0.240& 142.2& 2\\
 24 &   299.021832 &  -1.912362 & ------$\pm$ ------& 19.799$\pm$  0.187& 18.538$\pm$  0.095& 17.359$\pm$  0.061& 16.406$\pm$ 0.051&   7.47&  0.066&  65.2& 1\\
 25 &   299.015743 &  -1.914765 & ------$\pm$ ------& ------$\pm$ ------& 19.035$\pm$  0.150& 17.821$\pm$  0.093& 16.796$\pm$ 0.073&   5.28&  0.236&  31.6& 1\\
 26 &   298.991532 &  -1.914227 & 19.160$\pm$  0.063& 18.543$\pm$  0.060& 17.577$\pm$  0.040& 17.130$\pm$  0.049& 15.704$\pm$ 0.027&   9.17&  0.392& 103.3& 2\\
 28 &   298.952531 &  -1.297467 & ------$\pm$ ------& ------$\pm$ ------& ------$\pm$ ------& 18.445$\pm$  0.166& 17.431$\pm$ 0.133&   6.00&  0.545& 105.6& 2\\
 31 &   298.924510 &  -1.395874 & ------$\pm$ ------& ------$\pm$ ------& 19.665$\pm$  0.267& 18.506$\pm$  0.175& 17.576$\pm$ 0.150&   4.61&  0.284&  68.4& 2\\
 33 &   298.922812 &  -1.503068 & 19.927$\pm$  0.125& 19.060$\pm$  0.095& 18.333$\pm$  0.079& 17.272$\pm$  0.056& 16.360$\pm$ 0.049&   6.97&  0.137&  95.3& 1\\
 34 &   298.876502 &  -1.512565 & ------$\pm$ ------& ------$\pm$ ------& 17.991$\pm$  0.058& 16.693$\pm$  0.033& 15.694$\pm$ 0.027&   6.37&  0.064&  31.8& 2\\
\footnotetext[1]{The full version will be available on-line.}
\end{longtable}
\end{center}
\end{landscape}
\normalsize
\end{document}